\documentclass[12pt,indent]{article}
\usepackage{a4,latexsym,amsmath}
\usepackage[latin1]{inputenc}
\usepackage{float}
\usepackage{natbib}
\usepackage{graphics}
\usepackage[english]{babel}
\usepackage{tabularx}
\usepackage{lscape}
\usepackage{multirow}
\usepackage{rotating}
\usepackage{dsfont}
\usepackage{subfig}
\usepackage{bbm}
\usepackage[table]{xcolor}

\usepackage{amsmath}
\usepackage{amsfonts}

\usepackage{calc}
\usepackage{ifthen}

\usepackage{%
  longtable,
  array,
  booktabs,
  dcolumn,
  rotating,
  shortvrb,
  tabularx,
  units,
  multirow
}

\newenvironment{tabularsmall}
{ \footnotesize \sffamily \tabular } {
\endtabular
\normalfont }

\newcommand{\E}{\operatorname{E}}      
\newcommand{\var}{\operatorname{var}}
\newcommand{\cov}{\operatorname{cov}}

\providecommand{\norm}[1]{\lVert#1\rVert}

\newcommand{\betab}{{\boldsymbol{\beta}}}

\newcommand{\varepsilonb}{{\boldsymbol{\varepsilon}}}

\newcommand{\Hb}{\boldsymbol{H}}

\newcommand{\Ib}{\boldsymbol{I}}

\newcommand{\rb}{\boldsymbol{r}}

\newcommand{\Xb}{\boldsymbol{X}}
\newcommand{\xb}{\boldsymbol{x}}

\newcommand{\yb}{\boldsymbol{y}}

\newcommand{\blanco}[1]{}

\def\d{\displaystyle}

\usepackage{natbib}

\usepackage{setspace}

\begin{document}
\bibliographystyle{chicago}
\sloppy

\makeatletter
\renewcommand{\section}{\@startsection{section}{1}{\z@}%
        {-3.5ex \@plus -1ex \@minus -.2ex}%
        {1.5ex \@plus.2ex}%
        {\reset@font\Large\sffamily}}
\renewcommand{\subsection}{\@startsection{subsection}{1}{\z@}%
        {-3.25ex \@plus -1ex \@minus -.2ex}%
        {1.1ex \@plus.2ex}%
        {\reset@font\large\sffamily\flushleft}}
\renewcommand{\subsubsection}{\@startsection{subsubsection}{1}{\z@}%
        {-3.25ex \@plus -1ex \@minus -.2ex}%
        {1.1ex \@plus.2ex}%
        {\reset@font\normalsize\sffamily\flushleft}}
\makeatother



\newsavebox{\tempbox}
\newlength{\linelength}
\setlength{\linelength}{\linewidth-10mm} \makeatletter
\renewcommand{\@makecaption}[2]
{
  \renewcommand{\baselinestretch}{1.1} \normalsize\small
  \vspace{5mm}
  \sbox{\tempbox}{#1: #2}
  \ifthenelse{\lengthtest{\wd\tempbox>\linelength}}
  {\noindent\hspace*{4mm}\parbox{\linewidth-10mm}{\sc#1: \sl#2\par}}
  {\begin{center}\sc#1: \sl#2\par\end{center}}
}



\def\R{\mathchoice{ \hbox{${\rm I}\!{\rm R}$} }
                   { \hbox{${\rm I}\!{\rm R}$} }
                   { \hbox{$ \scriptstyle  {\rm I}\!{\rm R}$} }
                   { \hbox{$ \scriptscriptstyle  {\rm I}\!{\rm R}$} }  }

\def\N{\mathchoice{ \hbox{${\rm I}\!{\rm N}$} }
                   { \hbox{${\rm I}\!{\rm N}$} }
                   { \hbox{$ \scriptstyle  {\rm I}\!{\rm N}$} }
                   { \hbox{$ \scriptscriptstyle  {\rm I}\!{\rm N}$} }  }

\def\d{\displaystyle}

\title{ Probability and Non-Probability Samples:  Improving Regression Modeling by Using Data from Different Sources}
\author{Gerhard Tutz \\{\small Ludwig-Maximilians-Universit\"{a}t M\"{u}nchen}\\
 \small Akademiestra{\ss}e 1, 80799 M\"{u}nchen }


\maketitle

\begin{abstract}  
\noindent
Non-probability sampling, for example  in the form of online panels, has become a fast and cheap method to collect data. 
While reliable inference tools are available for classical probability samples, non-probability samples can yield strongly biased estimates  
since the selection mechanism is typically unknown. We propose a general method how to improve  statistical inference when in addition to a probability sample data from other sources, which have to be  considered  non-probability  samples,  are available.  
The method uses specifically tailored regression residuals  to enlarge the original data set by including observations from other sources that can be considered as stemming from the target population. Measures of accuracy of estimates are obtained by adapted  bootstrap techniques. It is demonstrated that the method can improve  estimates in a wide range of scenarios. For illustrative purposes, the proposed method is applied to two data sets. 

\end{abstract}


\newpage

\section{Introduction}

In the survey research literature there is an ongoing discussion on how to use  non-probability sample surveys to obtain
better estimates. Statistical analysis of non-probability
survey samples as, for example, web based surveys faces many challenges since the selection mechanism for
non-probability
samples is typically unknown and treating non-probability
samples as if they were a simple
random sample often leads to biased results \citep{baker2013summary,kim2021combining}.
Probability sampling theory provides a strong theoretical basis for the computation of the accuracy of estimates in probability samples.
For non-probability sample surveys no such framework is available. Typically, untested modeling assumptions have to be made that reflect 
how the sample differs from the population of interest \citep{cornesse2020review}. Forms of assumptions include quasi-randomization
or superpopulation modeling  \citep{jean1991theory,elliott2017inference}. One alternative strategy is based on sample matching. 
It is assumed that a probability sample is selected from the target population, the sample  must contain a set of auxiliary variables that is also available in the non-probability sample and can be used to match samples, see, for example, \citet{bethlehem2016solving}. A review on conceptual approaches  has been given by \citet{cornesse2020review}, see also \citet{kim2021combining}, \citet{wisniowski2020integrating}, 
\citet{elliott2017inference}, \citet{chen2020doubly}.
 
In survey methodology the focus typically is on the estimation of the population mean of the target population. Similar as in missing data problems assumptions on the probability to be included in the non-probability are made with covariates 
providing the required auxiliary information that is used to find weights in imputation estimators, see, for example, \citet{kim2021combining}.
The approach considered here is fundamentally different.
The focus is not on the mean of populations but on regression models. We investigate how the estimation of regression coefficients in a target population can be improved by including data from non-probability samples.  As in many approaches in survey methodology (see, for example,
\citet{vavreck20082006,lee2009estimation,brick2015compositional}) we assume that a probability sample of responses and covariates is available. The same covariates are observed in the non-probability sample, which is not representative of the target population due to the unknown sample selection mechanism. Diagnostic tools for regression models are used to decide if observations from the non-probability sample should be included in the analysis.

In Section \ref{sec:impr} first  residuals in regression are briefly sketched, then it is shown how they can be used to improve estimates of regression coefficients by enlarging the original data set. In Section \ref{sec:sim} some simulation results are given. In Section \ref{sec:stan} a bootstrap based concept how to obtain reliable estimates of standard errors is given. Some applications are found in Section \ref{sec:emp}.



\section{Improving Estimates}\label{sec:impr}

In the following it is assumed that  two samples are available:
\begin{itemize}
\item[]
$S_0=\{(y_i;\xb_i), i=1,\dots,n\}$ is a random sample from the target population, called the \textit{probability sample},   
\item[]
$S^{NP}=S_1 \cup S_2$ is a \textit{non-probability  sample} with $S_1=\{(y_i;\xb_i), i=1,\dots,n_1\}$ from the target population, and $S_2=\{(y_i;\xb_i), i=1,\dots,n_2\}$ being  a polluted sample.
\end{itemize}

In the non-probability sample it is not known which observations are from the target population and which are from the polluted sample. 
While the observations $S_1$ from the non-probability sample can be used to improve estimates of regression coefficients, inclusion of observations from sample  $S_2$ might yield distorted estimates since the distribution of responses and covariates in the population from which sample  $S_2$ is drawn might differ strongly from the distribution in the target population. The primary aim is to identify  which observations from $S^{NP}$ can be considered target population observations. It is basically a classification problem but without the learning sample, which is typically available in supervised classification problems. The role of the learning sample is taken by the probability sample but in a different way as in traditional classification problems as described, for example, in \citet{Bishop:06}.



\subsection{Regression Residuals  }

We first shortly sketch diagnostic tools in regression  that are useful to distinguish between target and non-target observations. More detailed descriptions of diagnostic tools are found in \citet{CooWei:82,FahTut:2001,fahrmeir2013regression}.

For observations $(y_i,\xb_i), i=1,\dots,n$ the linear model has the form $y_i=\xb^T_i \betab + \varepsilon_i$, where $\xb_i^T =(1,x_{i1},\dots x_{ip})$ is the vector of explanatory variables, $\betab^T=(\beta_0,\beta_1,\dots\beta_p)$ is the vector of coefficients, and the $\varepsilon_i$ are independent noise variables following a normal distribution $N(0,\sigma^2)$. In closed matrix form one has $\yb=\Xb\betab+\varepsilonb$, where $\yb^T=(y_1,\dots,y_n)$ is the vector of responses,
$\Xb$ is the design matrix  composed from the vectors of   explanatory   variables, and $\varepsilonb^T=(\varepsilon_1, \dots, \varepsilon_n)$
is the vector of errors. Fitting by least squares yields the familiar estimate $\hat\betab$.

Residuals provide  strong diagnostic tools. The simple residual $r_i =y_{i}-\xb_i^{T} \hat{\betab}$
measures the discrepancy between the actual observation and the f\/itted value $\hat y_{i} = \xb_i^{T}
\hat{\betab}$ and is a preliminary indicator for ill-f\/itting observations. However, it does not take the variability of the estimate
into account. It is simply derived that the vector of residuals $\rb =(r_1,\dots,r_n)^T = \yb - \hat\yb$, where $\hat\yb^T=(\xb_1^{T} \hat{\betab},\dots,\xb_n^{T} \hat{\betab})$
has covariance matrix $\cov(\rb)=\sigma^2 (\Ib-\Hb)$, where $\Hb$ is the projection matrix $\Hb = \Xb (\Xb^T \Xb)^{-1}
\Xb^T$.
Therefore, one obtains with the diagonal elements from $\Hb =
(h_{ij})$ the variance $\var(r_i) = \sigma^2/\sqrt{1-h_{ii}}$. Scaling of residuals to
the same variance produces the form
\[
\tilde r_i = \frac{r_i}{\sqrt{1-h_{ii}}},
\]
with $\var(\tilde r_i) = \sigma^2$. If, in addition, one divides by
the root of the estimated variance $\hat \sigma^2 = (\rb^T \rb) / (n-p-1)$,
where $p+1$ is the length of $\xb_i$, one obtains the {\em
studentized residual}
\[
r_i^* = \frac{\tilde r_i}{\hat \sigma} = \frac{y_i
-\hat\mu_i}{\hat\sigma \sqrt{1-h_{ii}}},
\]
which behaves much like a Student's $t$ random variable except for
the fact that the numerator and denominator are not independent.

An alternative useful approach is based on case deletion. Let
$\hat{\betab}_{(i)}$ denote the least-squared estimate resulting
from the  dataset in which observation $(y_i,\xb_i)$ is omitted.
The change in $\betab$ that results if the $i$-th observation is
omitted may be measured by
\[
\Delta_{i}\hat\betab=\hat\betab-\hat{\betab}_{(i)}.
\]
For the linear model no refitting is needed since an explicit form is available given by $\Delta_{i}\hat\betab=(\Xb^T\Xb)^{-1}\xb_{i}r_{i}/(1-h_{ii})$, see, for example, \citet{CooWei:82}.

\subsection{Improved Estimates by Enlarging the Data Set}

The basic concept is to use  diagnostic tools for regression models to check if   observations from the non-probability sample should be considered as data from the target population or not. The main source of information is the   probability sample, which is known to be from the target population.
For each single observation from the non-probability sample it is investigated if it is an ``outlier''.

\medskip
The basic algorithm is the following: 

\begin{itemize} 
\item[(1)] Fit the regression model  $y_i = \xb_i^T\betab+ \varepsilon_i$ by using the  probability sample to obtain $\hat\betab_0$.
\item[(2)] Fit the regression model to data $S_0 \cup \{(y_i,\xb_i)\}$, where $(y_i,\xb_i) \in S^{NP}$, that is, one observation from the non-probability sample is added to the probability sample, to obtain $\hat\betab_{(i)}$ and the corresponding studentized residual $r^*_{s,i}$               
\item[(3)]
Observation $(y_i,\xb_i)$ is considered to be from the target population if two criteria are fulfilled:
\[
|r^*_{s,i}| \le t_s, 
\]
where $t_s$ is a threshold for the studentized residuals, and

\[
\Delta_{i}=\frac{\norm{\hat \betab_0 - \hat \betab_{i}}}{\norm{\hat \betab_0}} < t_c, 
\]
where $t_c$ is a threshold for the change in parameters when observation $(y_i,\xb_i)$ is included. 
\item[(4)] All observations that fulfill both criteria are added to  the probability sample to obtain the \textit{extended sample}, which is used to compute the final estimate.
\end{itemize}

The first criterion uses the studentized residual in the enlarged sample $S_0 \cup \{(y_i,\xb_i)\}$. If the observation $(y_i,\xb_i)$ is from the target population its studentized residual should not be too large. The second criterion is based on case deletion  but case deletion in the enlarged sample  
$S_0 \cup \{(y_i,\xb_i)\}$. It compares parameter estimates in the probability sample with estimates in the  sample enlarged by one observation. Thus it is  case deletion regarding the enlarged sample but can also be seen as case addition regarding the original probability sample. When adding the $i$-th observation the change in parameters should not be too large   if it is from the target population. 


\subsubsection*{Choice of Thresholds}

The procedure  depends on the choice of the thresholds $t_s$ and $t_c$, which we link  to quantiles. Given a nominal error level $\alpha_{st}$, we choose \textit{for  the studentized residuals} $t_s=q_{1-\alpha_{st}}$, where $q_{1-\alpha_{st}}$ denotes the $(1-\alpha_{st})$- quantile of the standard normal distribution,   in the applications we use $\alpha_{st}=0.05$. 

The choice of the threshold \textit{for the change in parameters} is based on  the distribution of changes in parameters that are to be expected in a sample from the target population.
In the probability  sample, which is from the target population, there is natural variation in estimates if one observation is omitted. We explicitly investigate this  distribution in the following way:
\begin{itemize} 
\item[]
(a) For all observations from the probability sample fit the  model   by deleting one observation at a time. Fitting the model by using observations $S_0$ without $\{(y_i,\xb_i)\}$, that is, without the $i$-th observation , yields $\hat\betab_{-i}$
\item[] 
(b) Consider the  distribution of changes in parameters if one observation is omitted with changes given by
\[
ch_i =\frac{\norm{\hat \betab_0 - \hat \betab_{-i}}}{\norm{\hat \betab_0}}. 
\]
\item[] 
(c) Choose  $t_c$ as the $(1-\alpha_{ch})$-quantile  of the \textit{empirical} distribution of the values of $ch_i$.
\end{itemize}

The distribution of $ch_i$ shows which  changes are to be expected if observations are from the target population. If the relative change when including an observation from the non-probability sample is below the threshold $t_c$ it is considered as coming from the target population.

The method to enlarge the data set uses two tuning parameters $\alpha_{st}$ and  $\alpha_{ch}$. If $\alpha_{st}$ and  $\alpha_{ch}$ are small many observations from the non-probability sample are included in the extended data set, if they are large the number of included observations is small. 
Thus, the  $\alpha$-values are an indicator for the size of the extended sample, with increasing $\alpha$-values the extended sample gets smaller and in the extreme case consists of the original probability sample.

It should be noted that the $\alpha$-values are not  significance levels in the traditional sense. They are tuning parameters, but it seems quite natural to describe them in distributional terms. As tuning parameters they can also be chosen in a data-driven way, for example, by cross-validation. For simplicity one can also use $\alpha_{st}=\alpha_{ch}$.

\subsubsection*{Cross-Validation}
The tuning parameters $\alpha_{st}$ and  $\alpha_{ch}$ can be chosen as fixed values. An alternative is to choose them in a data driven way by cross-validation. Then the probability sample is split randomly into $k$ distinct data sets $S_{01},\dots,S_{0k}$ of approximately equal size. 
One considers $S_0 \setminus S_{0j}$ as learning data set, which is used to obtain an enlarged sample, and $S_{0j}$ as validation sample, for which, based on the extended sample, predicted values are computed. The discrepancy between the predicted values and the observed values in  
$S_{0j}$ shows how well the procedure works in terms of prediction. Computing the discrepancy  for all sub samples $S_{0j}$ and averaging yields a measure for the performance of fixed tuning parameters $\alpha_{st}$ and  $\alpha_{ch}$. Computation of the discrepancies on a grid of $\alpha$ 
values allows to choose the values which show best performance. As discrepancy measure one can use the squared error and $\alpha$ values from a grid of values. One can compute on the whole grid or use a less demanding cross-validation method that uses a reduced grid of values with $\alpha_{st}=\alpha_{ch}$. 
 
\subsection{Accuracy Measures }

For the parameter estimate $\hat\betab$ a measure of accuracy is the mean squared error $MSE=||\hat\betab-\betab||^2$. It is used to compare the estimates obtained for the probability sample and the extended sample. One can also compare the estimates of the two methods directly. 
Let $\hat\betab_P$ denote the parameter estimate when using the probability sample, and $\hat\betab_E$ the estimate obtained from the extended sample. A measure that compares the accuracy of these estimates 
is the relative mean squared error,
\[
MSE_r = \frac{||\hat\betab_{P}-\betab||^2}{||\hat\betab_{ext}-\betab||^2}. 
\]
It indicates how much larger the mean squared error of the simple probability sample estimator is than the estimator obtained in the extended data set.
It can also be computed for the reduced coefficient vector that contains the coefficients on explanatory variables only. This seems sensible if one is interested in the impact of covariates only. If, however, one aims at prediction the whole vector of coefficients makes more sense. The choice is of minor importance since we found that the improvement obtained by the enlargement of the sample is very similar in both types of parameter vectors.

Further measures of performance refer to the identification of observations in the non-probability sample. Similar as in classification problems one can consider 
\begin{itemize} 
\item[] hits: the proportion of observations correctly identified as coming from the target population,
\item[] false positives: the proportion of observations incorrectly  identified as coming from the target population.
\end{itemize}

A large number of hits means that the information in the non-probability sample is efficiently used, whereas a  large number of false positives is an indicator that estimates might be distorted. In the ideal case all the information in the non-probability sample is used but no distortion of estimates is observed. It should be mentioned that inclusion of observations from the polluted sample not necessarily entails biased estimates since only observations are included that do not change the parameters strongly.

\section{Some Simulations }\label{sec:sim}
\subsection{One Predictor }

For illustration we consider briefly the case of just one predictor with fixed parameters. In a small simulation study we compare  the estimators based on the probability sample and the extended sample.  The simulation uses $n=40$, $n_1=200$, $n_2=200$. The distribution of covariates and the 
parameters are specified in the following way.
\begin{itemize} 
\item[]Setting (a):
Probability sample: normally distributed,  $\mu_0=\E(x)=1$,
regression parameters: $\betab^T=(\beta_{0},\beta_{1})=(1,1)$,  $\var(\varepsilon)=1$

Polluting sample: normally distributed, $\E(x)=2$, 
regression parameters: $\betab^T=(2,-1)$,  $\var(\varepsilon)=4$

\item[]Setting (b):
Probability sample: normally distributed,  $\mu_0^T=\E(x)=1$,
regression parameters: $\betab^T=(\beta_{0},\beta_{1})=(1,1)$,  $\var(\varepsilon)=1$

Polluting sample: normally distributed, $\E(x)=2$, 
regression parameters: $\betab^T=(3,0.5)$,  $\var(\varepsilon)=9$


\end{itemize}

\begin{figure}[H]
	\centering
		\includegraphics[width=0.45\textwidth]{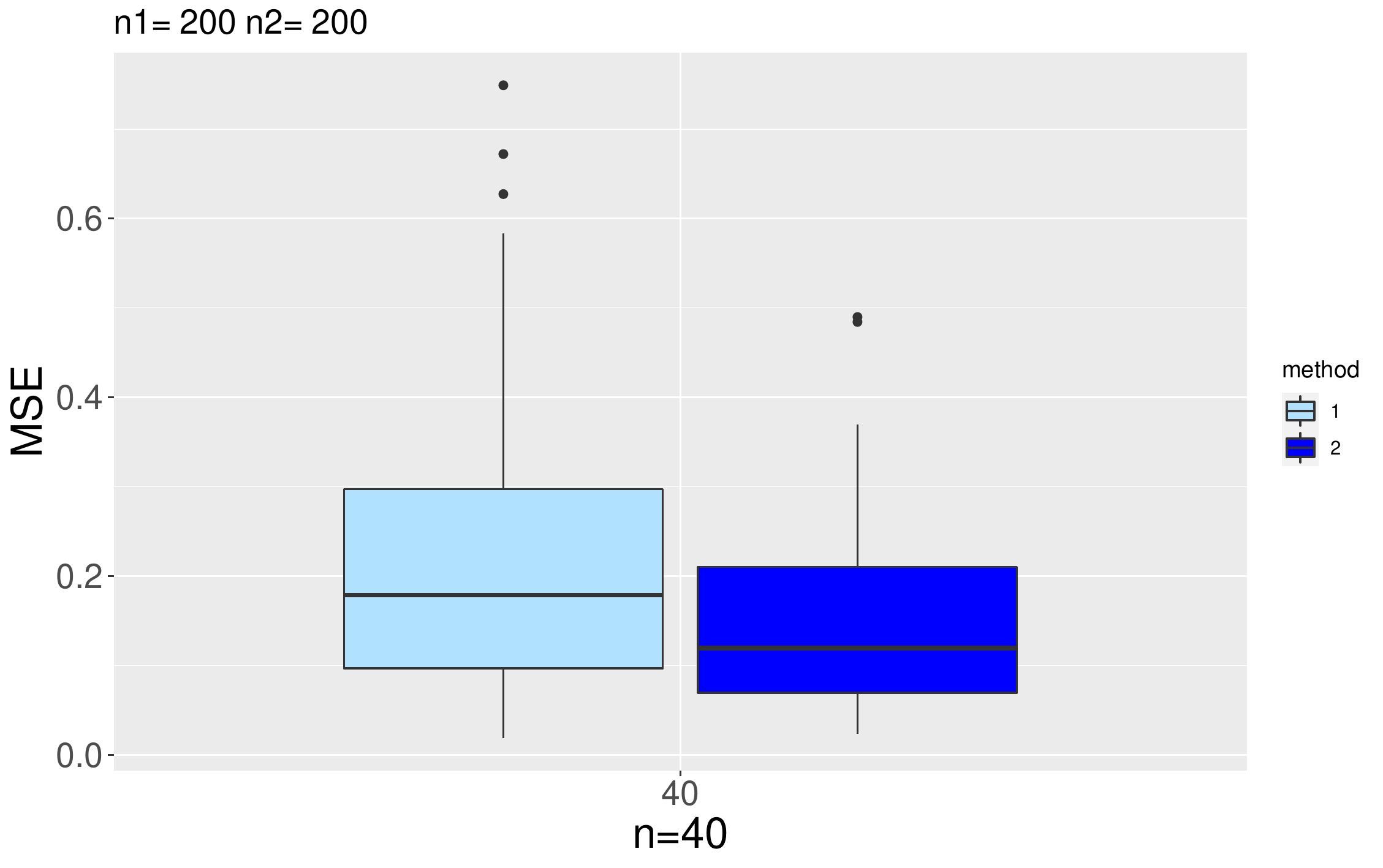}
  \includegraphics[width=0.45\textwidth]{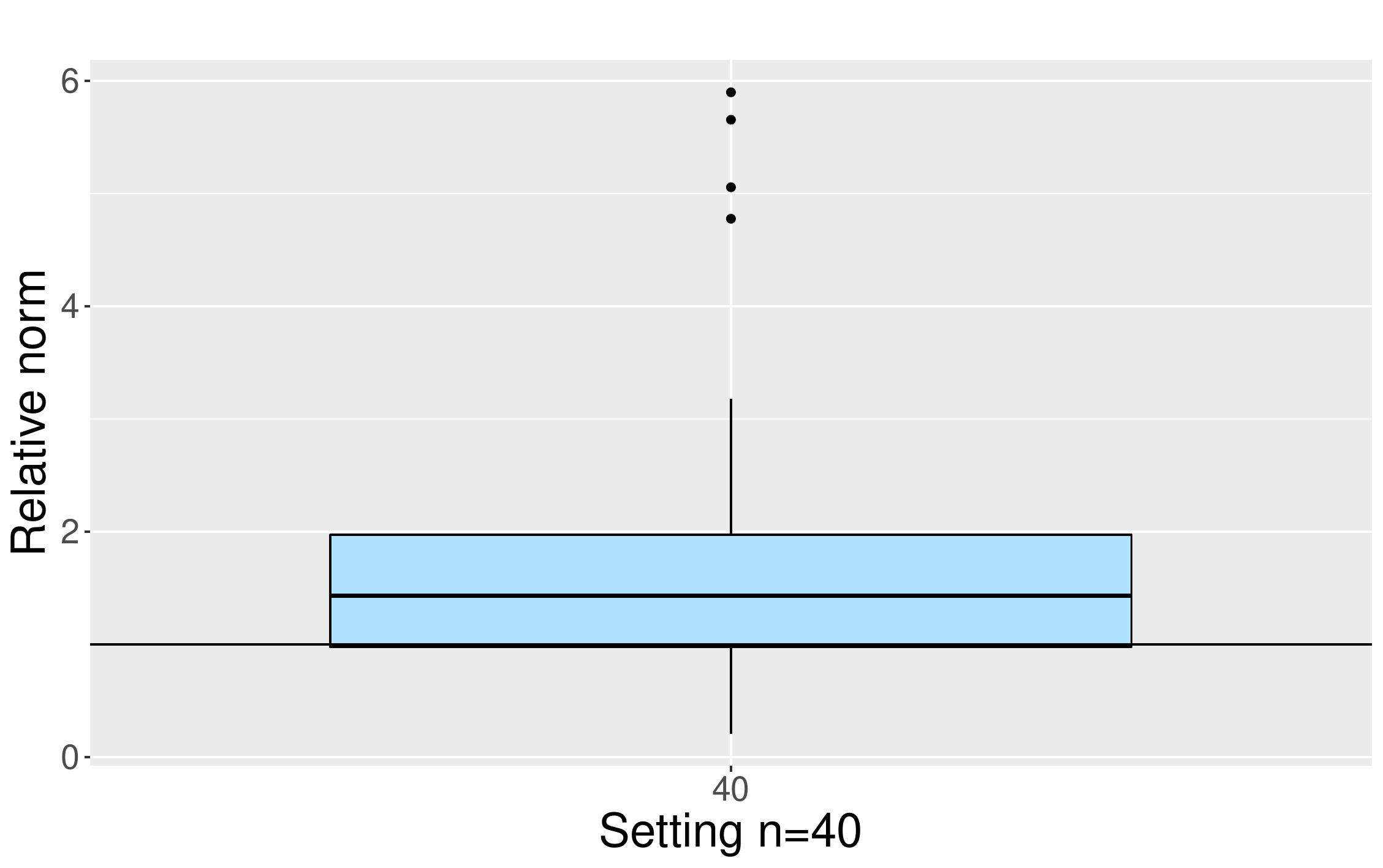}
  \includegraphics[width=0.45\textwidth]{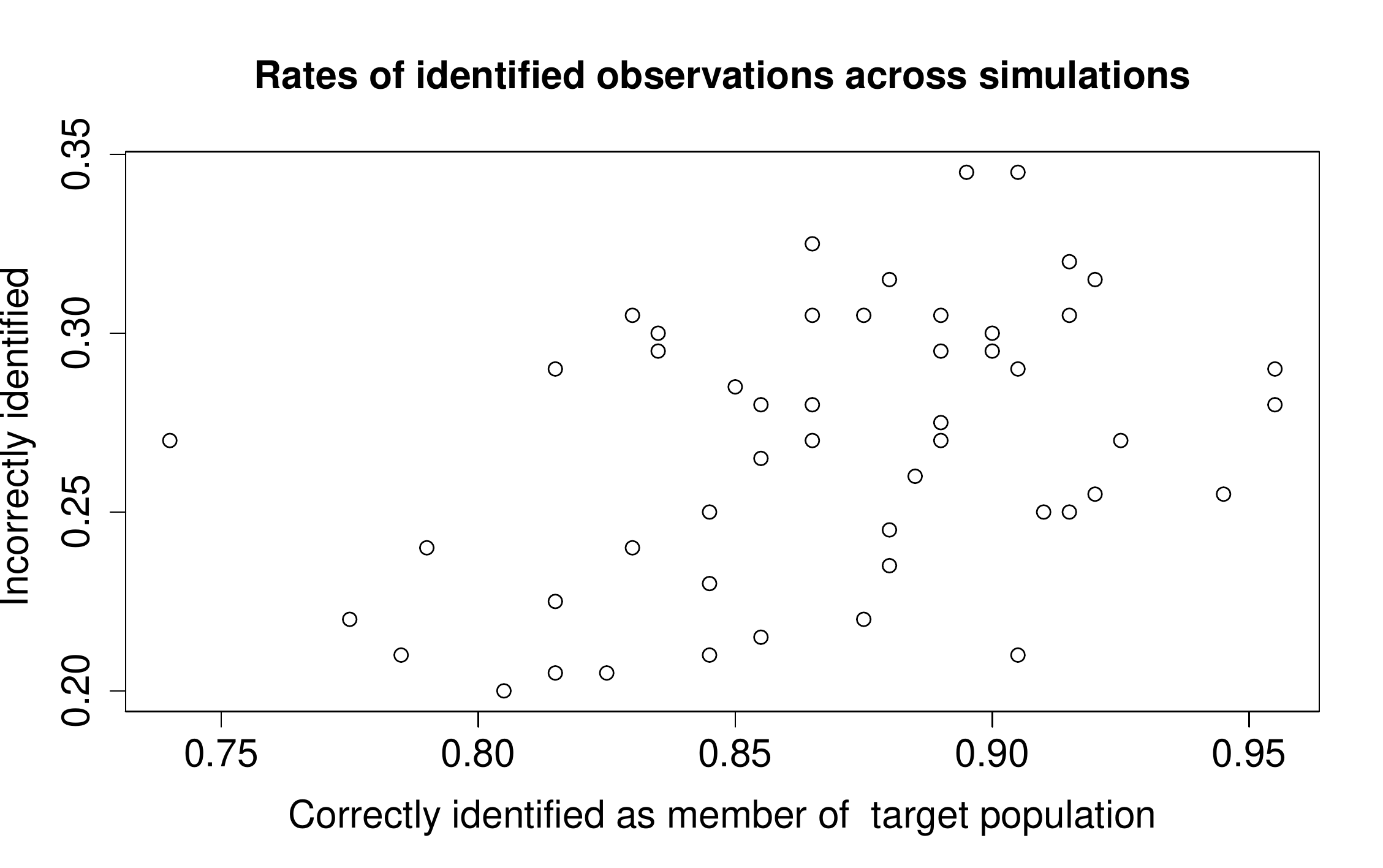}
  \includegraphics[width=0.45\textwidth]{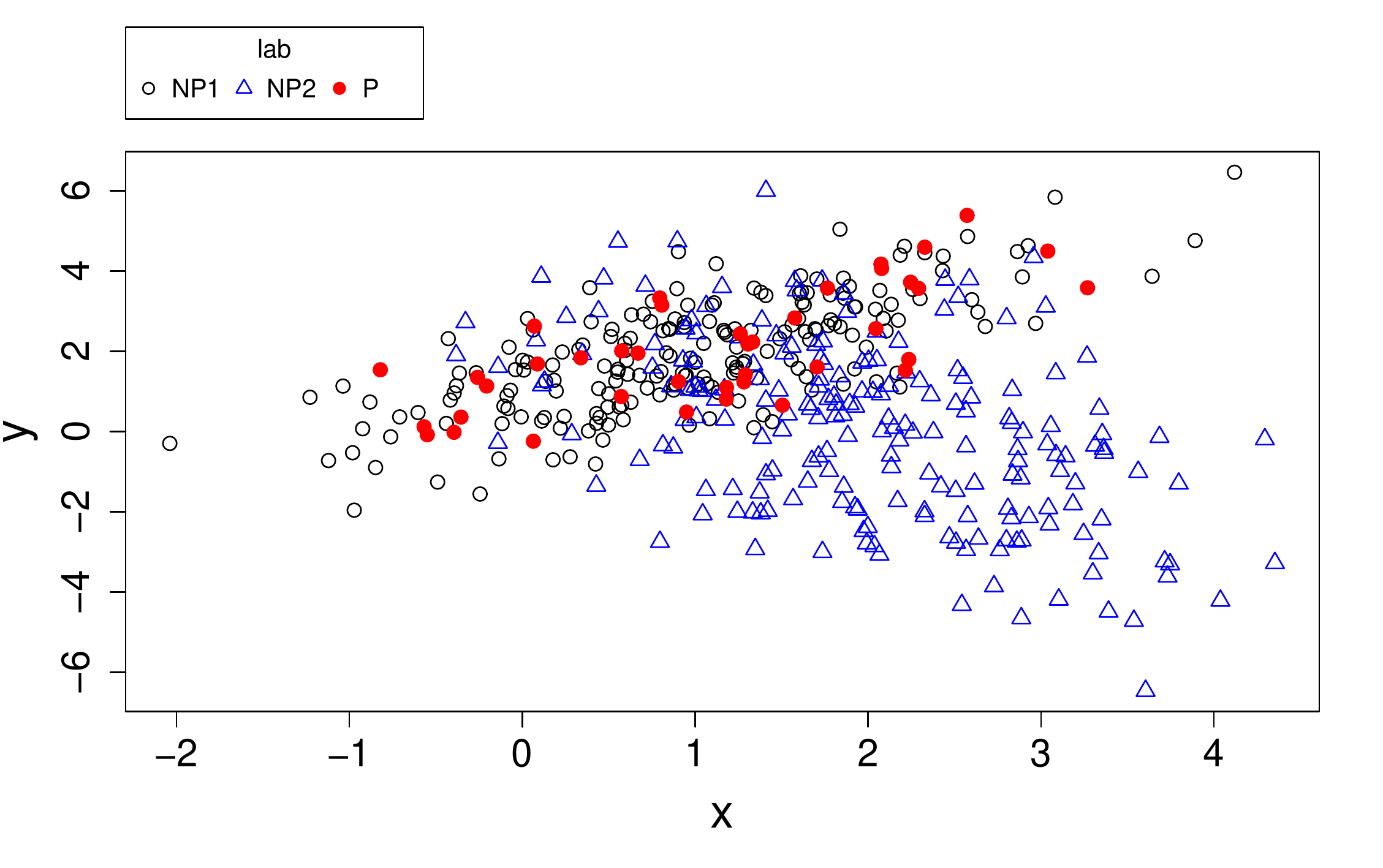}
  \includegraphics[width=0.45\textwidth]{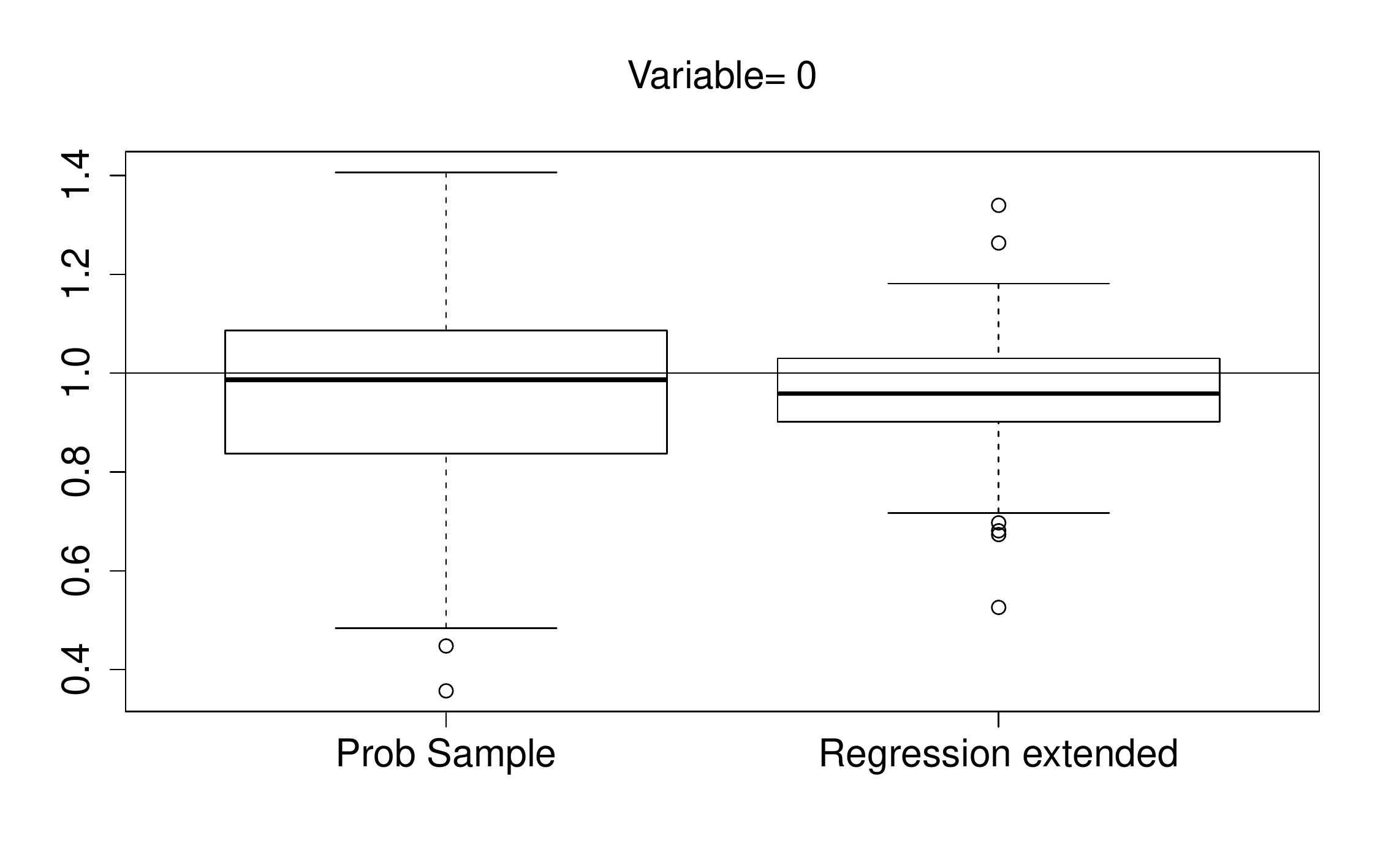}
  \includegraphics[width=0.45\textwidth]{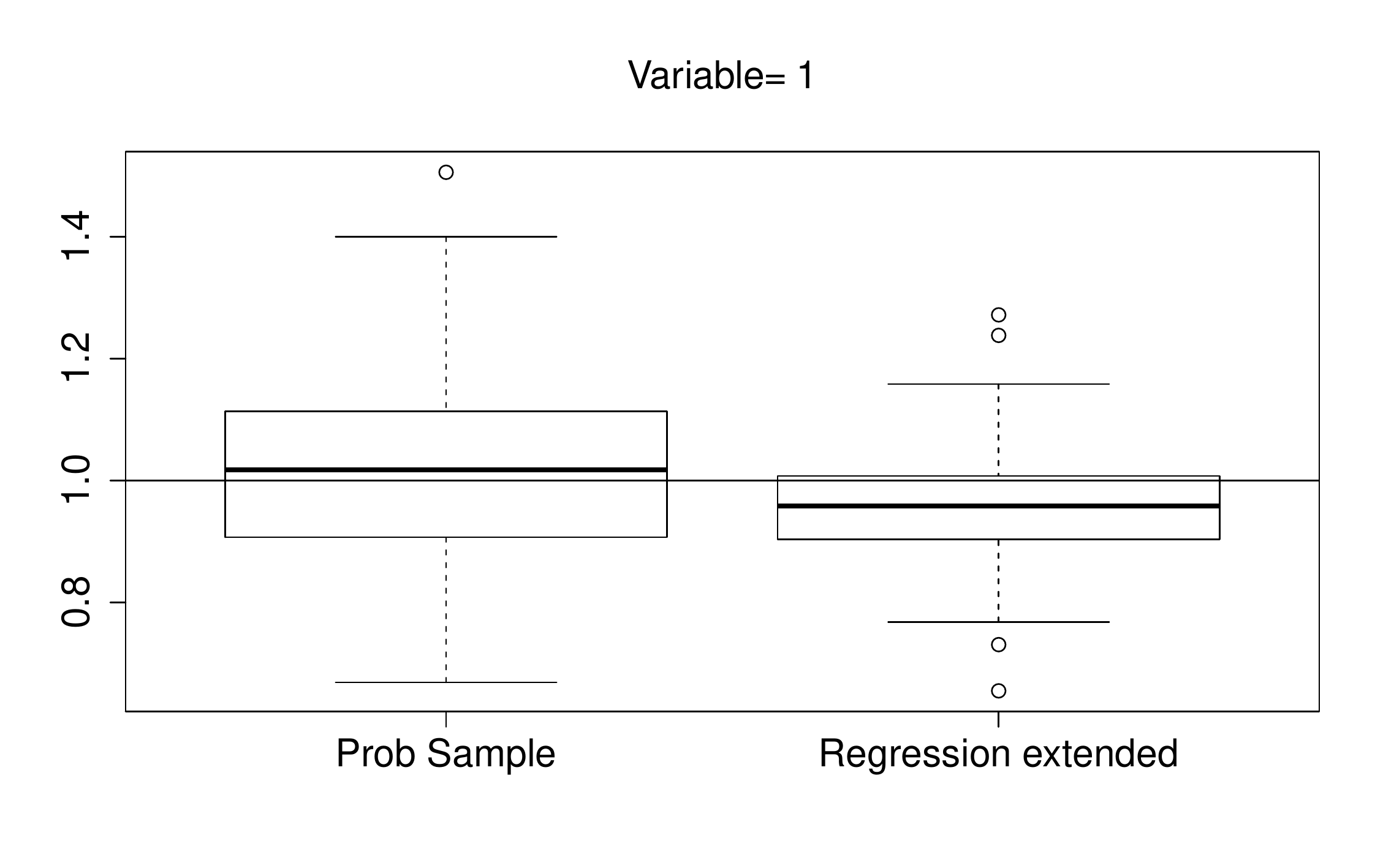}  
	 \caption{Simulation univariate predictor (setting (a)); upper left picture:  MSEs for the probability sample estimator (light) and the estimator based on the extended sample (dark); upper right picture: relative mean squared errors; second row, left: proportions of correctly and incorrectly identified observations; second row, right: observations  for the first simulated data set; third row: box plots of the estimates for the two parameters $\beta_0$ and $\beta_0$ with the horizontal line indicating the true value. }
  \label{fig:sima}
\end{figure}

Figure \ref{fig:sima} shows the results  for $\alpha_{st}=\alpha_{ch}=0.05$ (setting (a)). The upper left picture shows the MSEs for the probability sample estimator (PSE, light) and the estimator based on the extended sample (ExtE, dark). It is seen that the MSE of the PSE (light blue) is distinctly smaller than the MSE of the ExtE (dark blue). This is also seen from the upper right picture which shows the relative mean squared errors. The inclusion of observations from the non-probability sample yields strongly improved estimates. The second row shows the proportions of correctly and incorrectly identified observations in the 50 simulated data sets (left) and the observations of both variables for the first simulated data set (right). It is seen that the number of observations from the non-probability sample that are correctly identified as coming from the target population is always large, the number of incorrectly assumed as coming from the target population is between 0.2 and .35. But as the distribution of the response and the predictor in the right picture shows incorrectly identified observations do not necessarily affect the accuracy of the estimator, it is only essential to exclude all observations that might include bias. The last row shows the box plots of the estimates for the two parameters $\beta_0$ and $\beta_0$ with the horizontal line indicating the true value. It is seen that in particular the variability of estimates is smaller for the extended sample denoted by regression extended. 

\begin{figure}[H]
	\centering
		\includegraphics[width=0.45\textwidth]{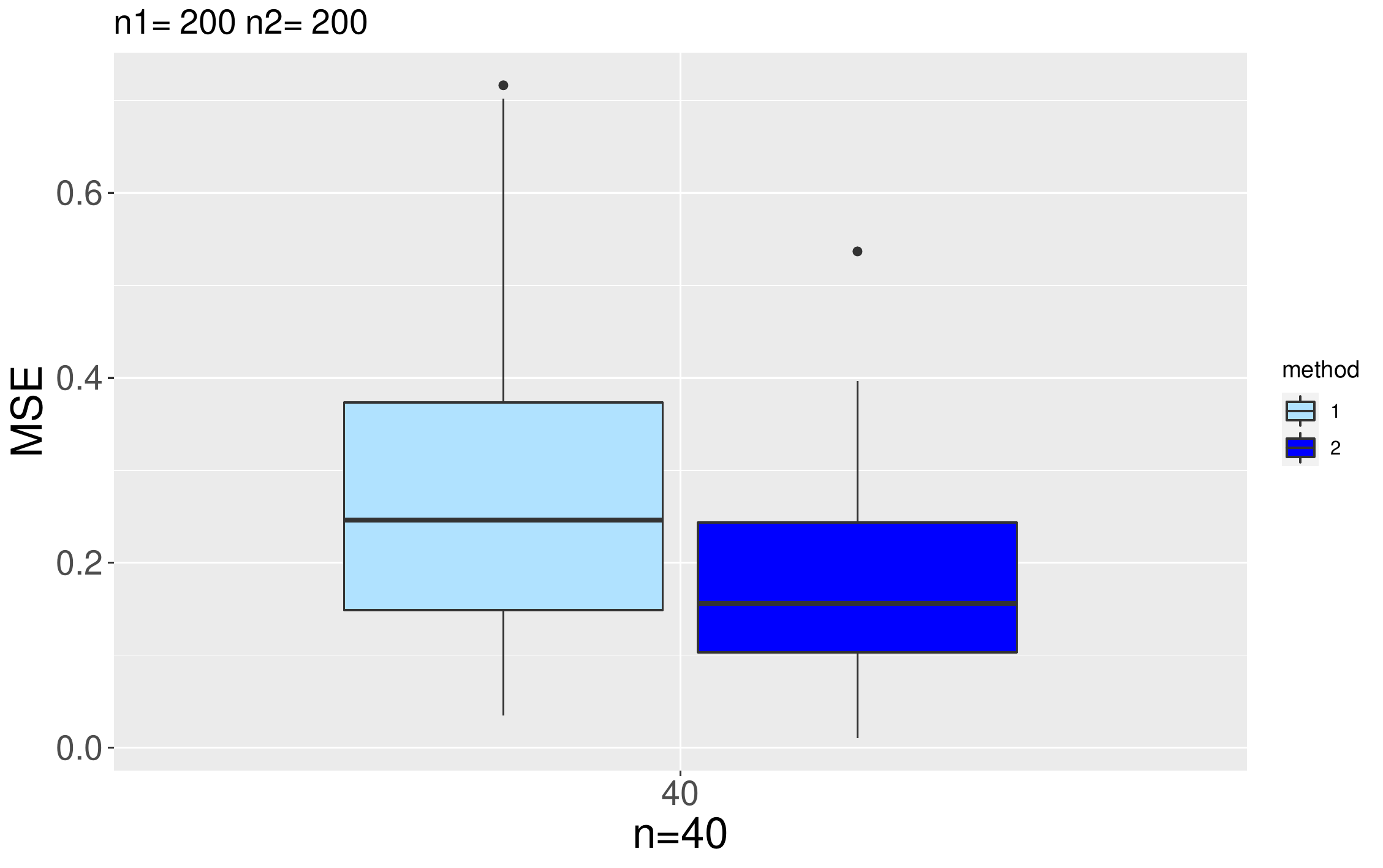}
  \includegraphics[width=0.45\textwidth]{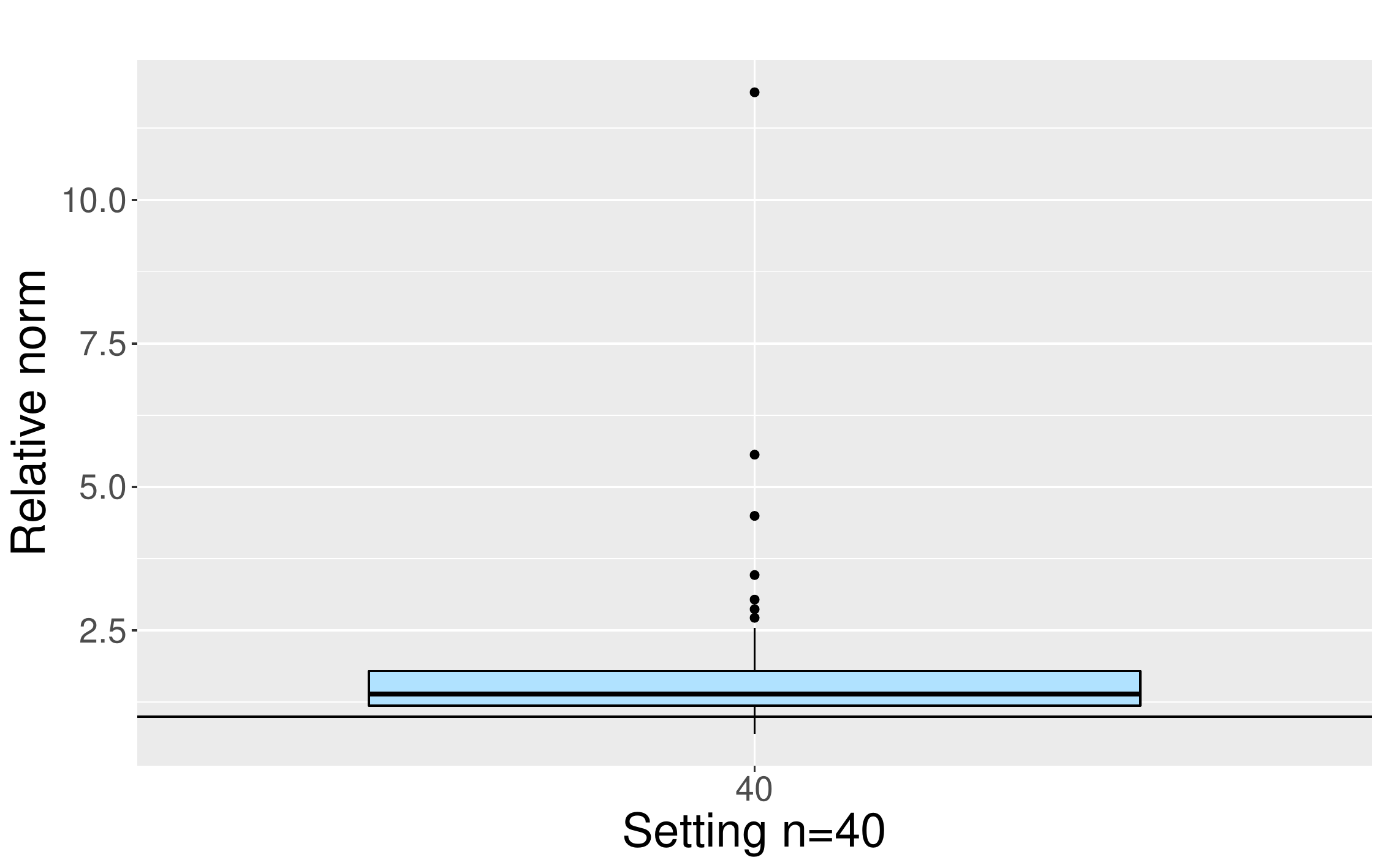}
  \includegraphics[width=0.45\textwidth]{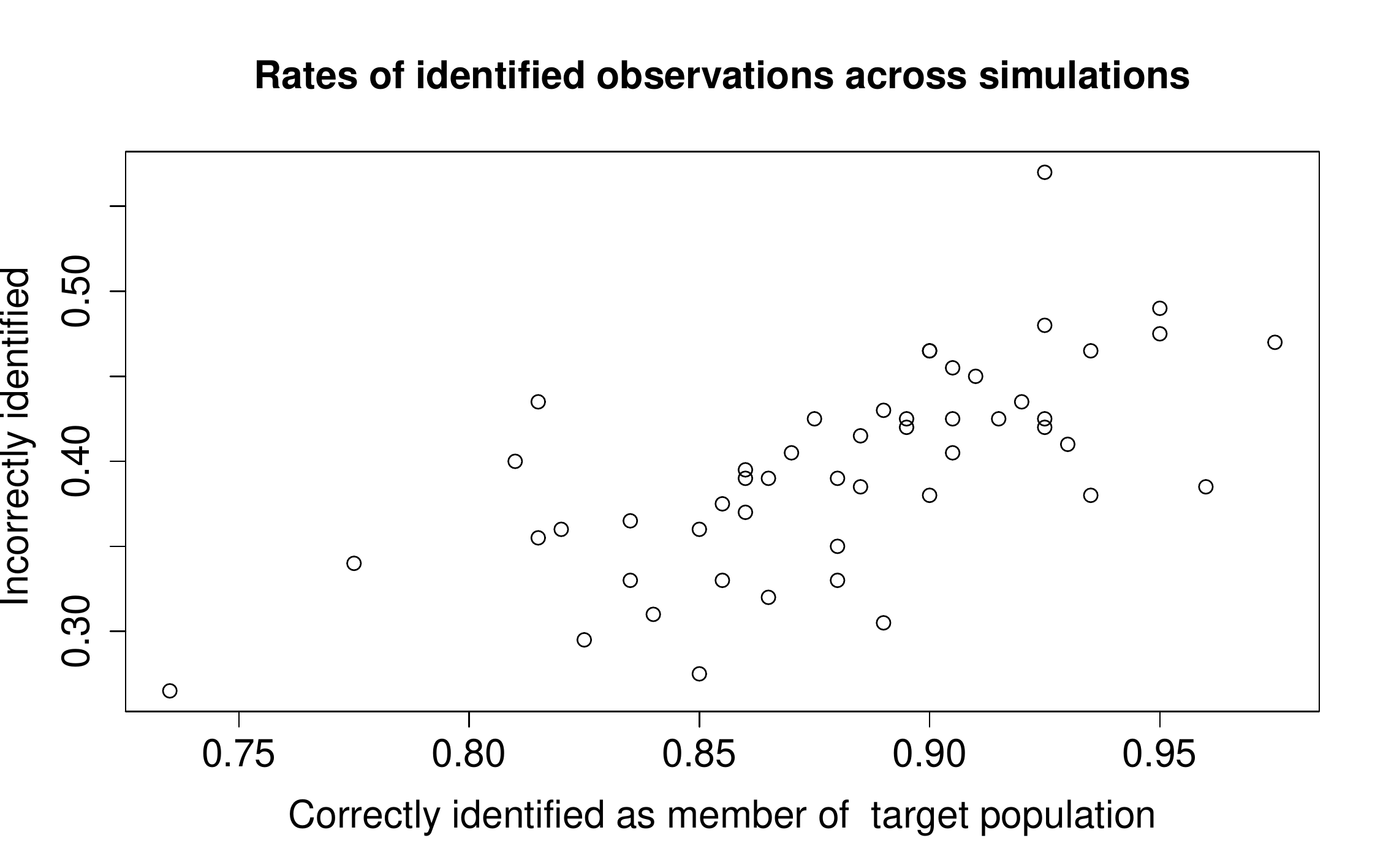}
  \includegraphics[width=0.45\textwidth]{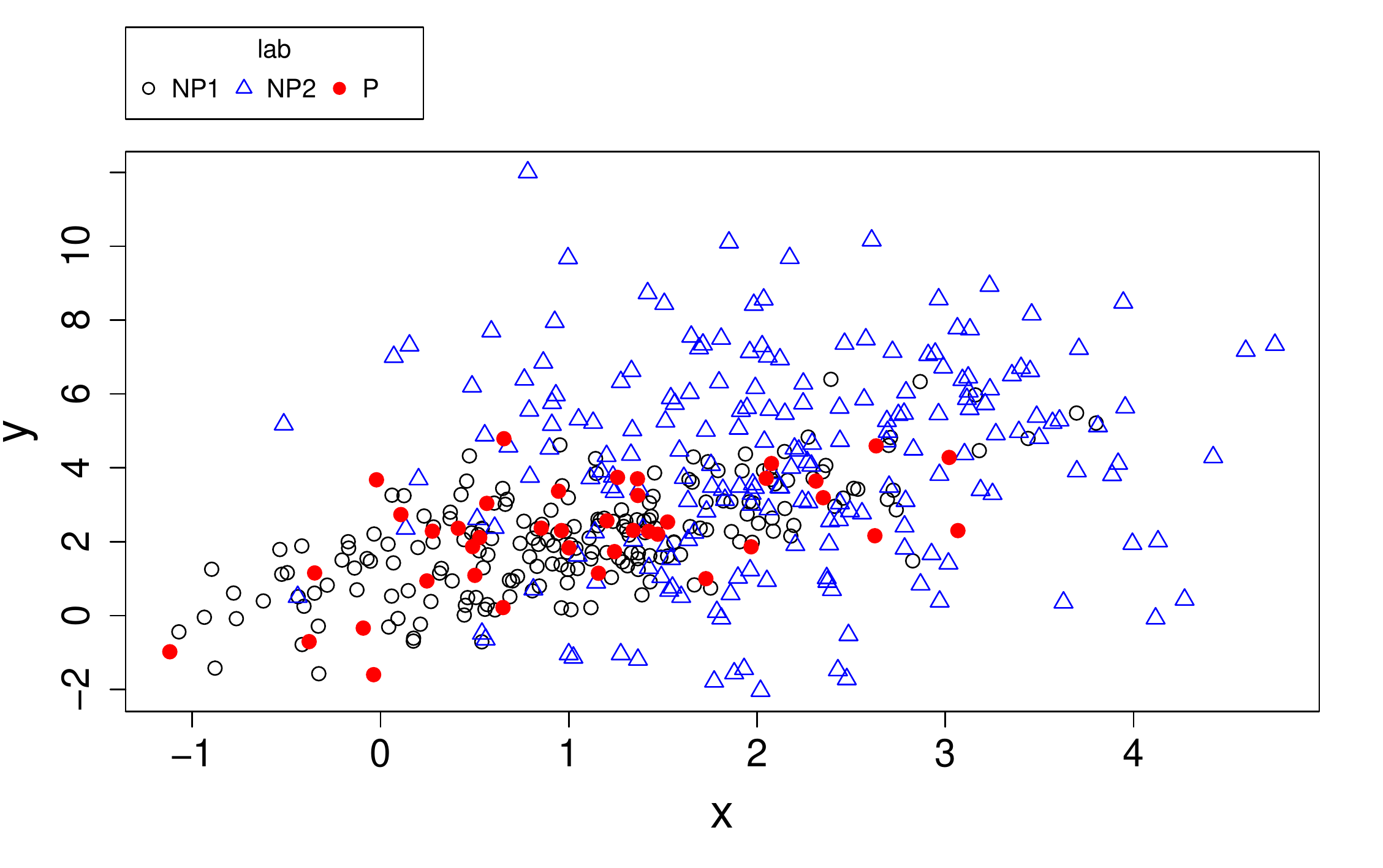}
  \includegraphics[width=0.45\textwidth]{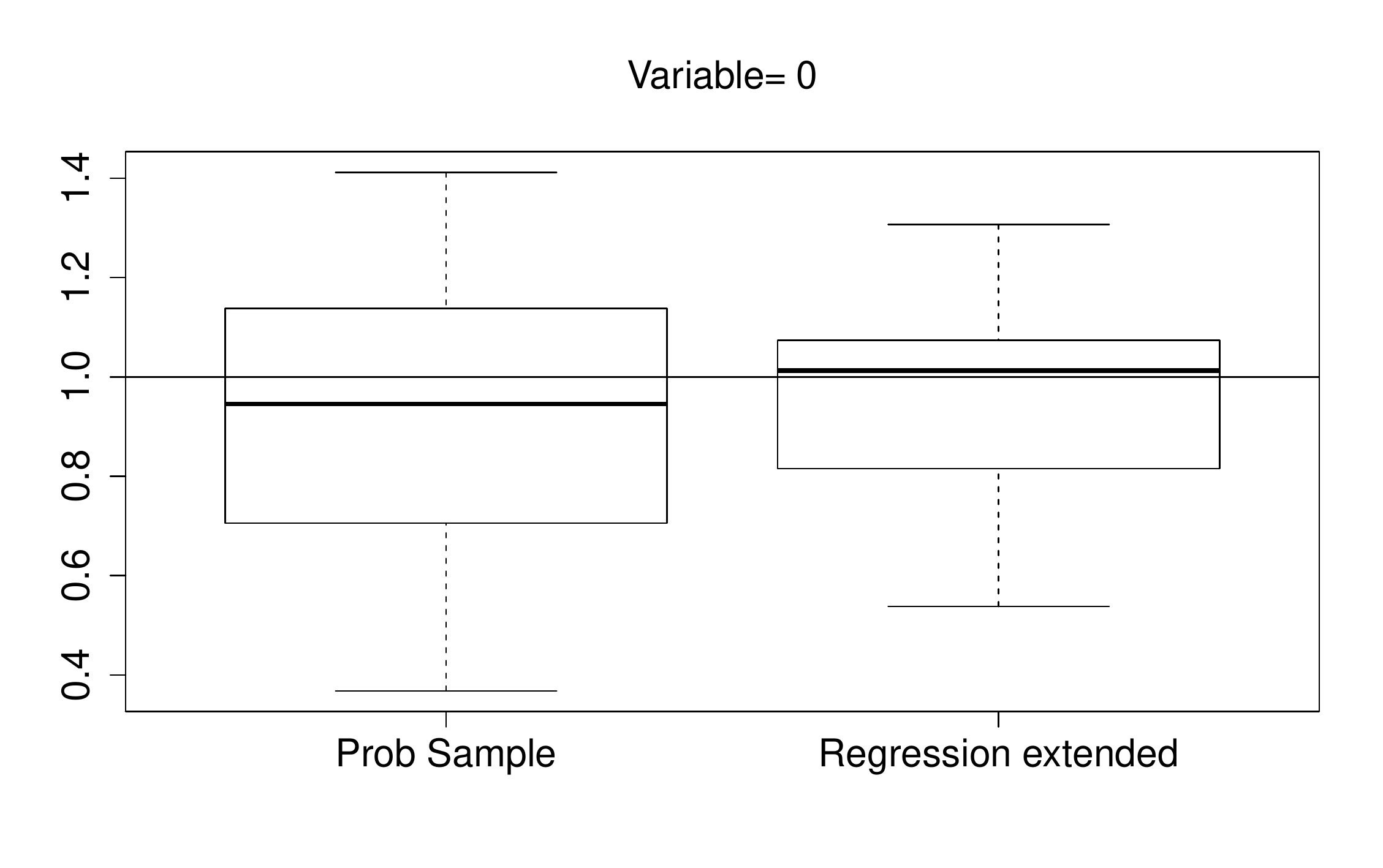}
  \includegraphics[width=0.45\textwidth]{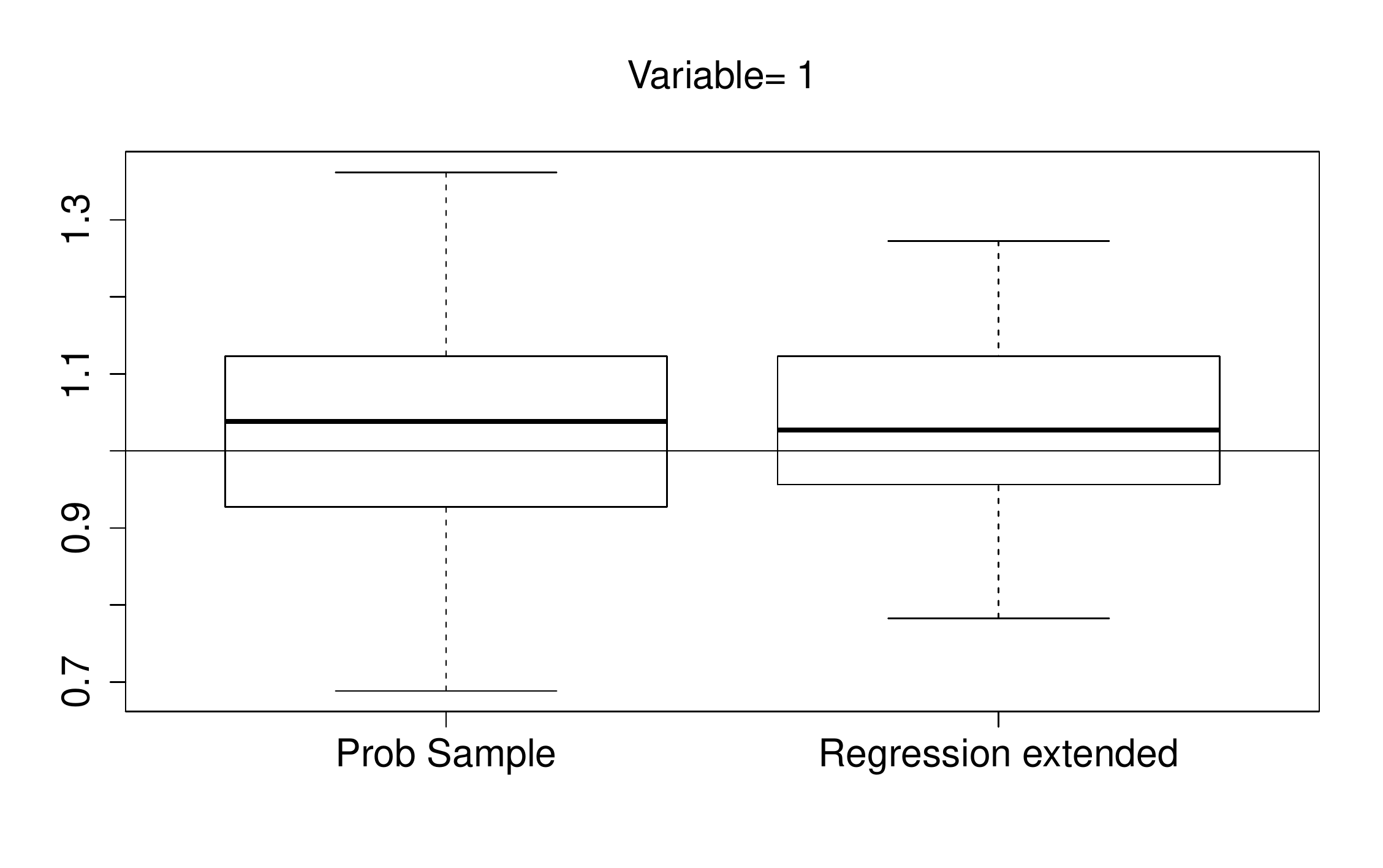}  
	 \caption{Simulation univariate predictor (setting (b)); upper left picture:  MSEs for the probability sample estimator (light) and the estimator based on the extended sample (dark); upper right picture: relative mean squared errors; second row, left: proportions of correctly and incorrectly identified observations; second row, right: observations  for the first simulated data set; third row: box plots of the estimates for the two parameters $\beta_0$ and $\beta_0$ with the horizontal line indicating the true value.}
  \label{fig:simb}
\end{figure}

Figure \ref{fig:simb} shows the corresponding pictures for setting (b). Although the number of incorrectly identified observations is much larger than in setting (a) the improvement in terms of MSE is very strong. The reason is that the observations that are incorrectly included do not distort the estimate since the slope in the polluted sample  differes from the slope in the target population but not as strongly as in setting (a), where the signs of the slopes differ.
For further illustrations see Appendix.

\subsection{Multivariate Predictors } 

In the following we show the results of selected simulations for more than one explanatory variables (100 data sets). Regression parameters and location of the explanatory variables in the polluted sample are not fixed  but vary across data sets. The location of explanatory variables in the probability sample is determined by the mean vector $\mu_0^T=\E(\xb)^T=(\mu_{00},\dots,\mu_{0p})$. In the polluted sample to each component a normally distributed shift is added, $\E(x_i)=\mu_{0i}+\sigma_{loc}*N(0,1)$, with $\sigma_{loc}$ determining the strength of the shift. 
In the same way the regression parameters in the polluted sample are determined as deviations from the parameters in the probability sample, $\beta_{i}=\beta_{0i}+\sigma_{par}*N(0,1)$,  with $\sigma_{par}$ determining the strength of the shift. We consider the following settings.

\begin{itemize} 
\item[]Setting 1: Four predictors
 
Probability sample: normally distributed, predictors pairwise correlated with correlation coefficient 0.3, $\mu_0^T=\E(\xb)=(1,1,1,1)$,
regression parameters: $\betab^T=(\beta_{0},\dots,\beta_{4})=(1,1,2,3,4)$,  $\var(\varepsilon)=1$

Polluting sample: normally distributed, predictors pairwise correlated with correlation coefficient 0.3, $\E(x_i)=\mu_{0i}+\sigma_{loc}*N(0,1)$, $\sigma_{loc}$=1,
regression parameters: $\beta_{i}=\beta_{0i}+\sigma_{par}*N(0,1)$, $\sigma_{par}=1$, $\var(\varepsilon)=2$

\item[]Setting 2: Same as setting 1 but now stronger variation in polluting sample $\sigma_{par}=2$ or $\sigma_{par}=4$

\item[]Setting 3: Eight predictors

Probability sample:normally distributed, predictors pairwise correlated with correlation coefficient 0.3, $\mu_0^T=\E(\xb)=(1,\dots,1)$,
regression parameters: $\betab^T=(\beta_{0},\dots,\beta_{8})=(1.0, 0.5, 1.0, 1.5, 2.0, 2.5, 3.0, 3.5, 4.0)$,  $\var(\varepsilon)=1$

Polluting sample: normally distributed, predictors pairwise correlated with correlation coefficient 0.3, $\E(x_i)=\mu_{0i}+\sigma_{loc}*N(0,1)$, $\sigma_{loc}$=2,
regression parameters: $\betab=(1.0, 0.5, 1.0, 1.5, 2.0, 2.5, 3.0, 3.5, 4.0)$, $\sigma_{par}=1$, $\var(\varepsilon)=4$

\end{itemize}

\begin{figure}[H]
	\centering
		\includegraphics[width=0.45\textwidth]{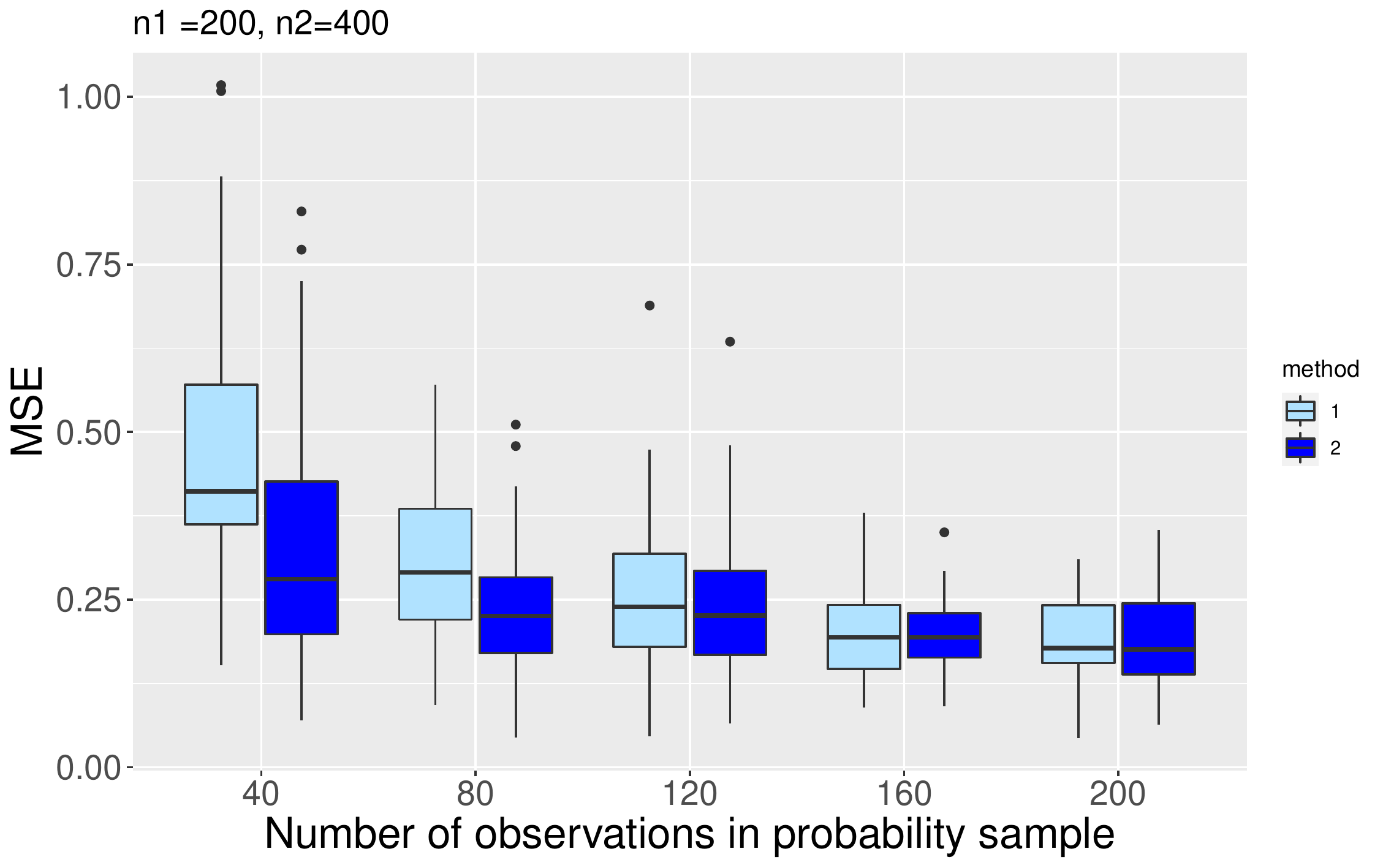}
  \includegraphics[width=0.45\textwidth]{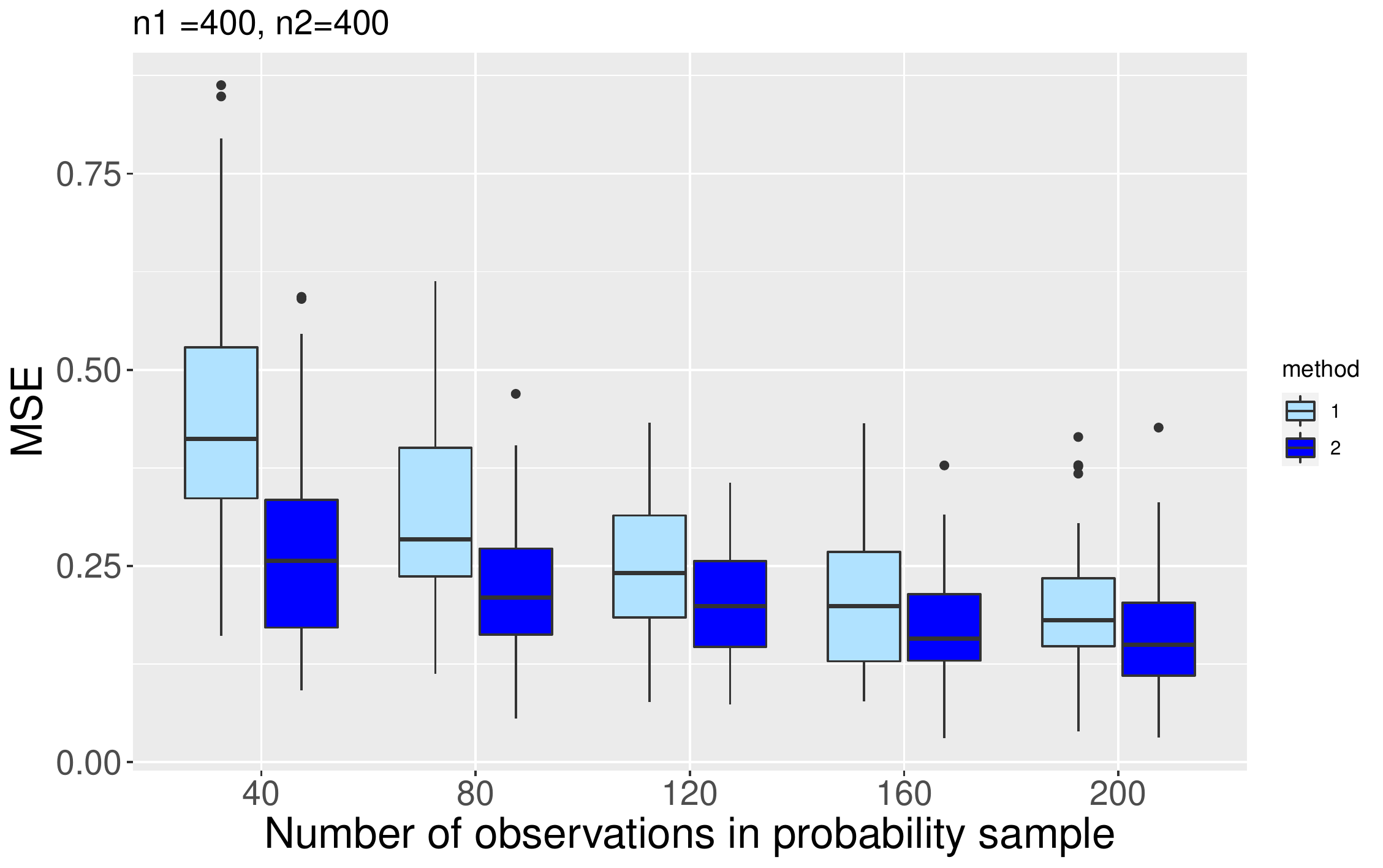}
  \includegraphics[width=0.45\textwidth]{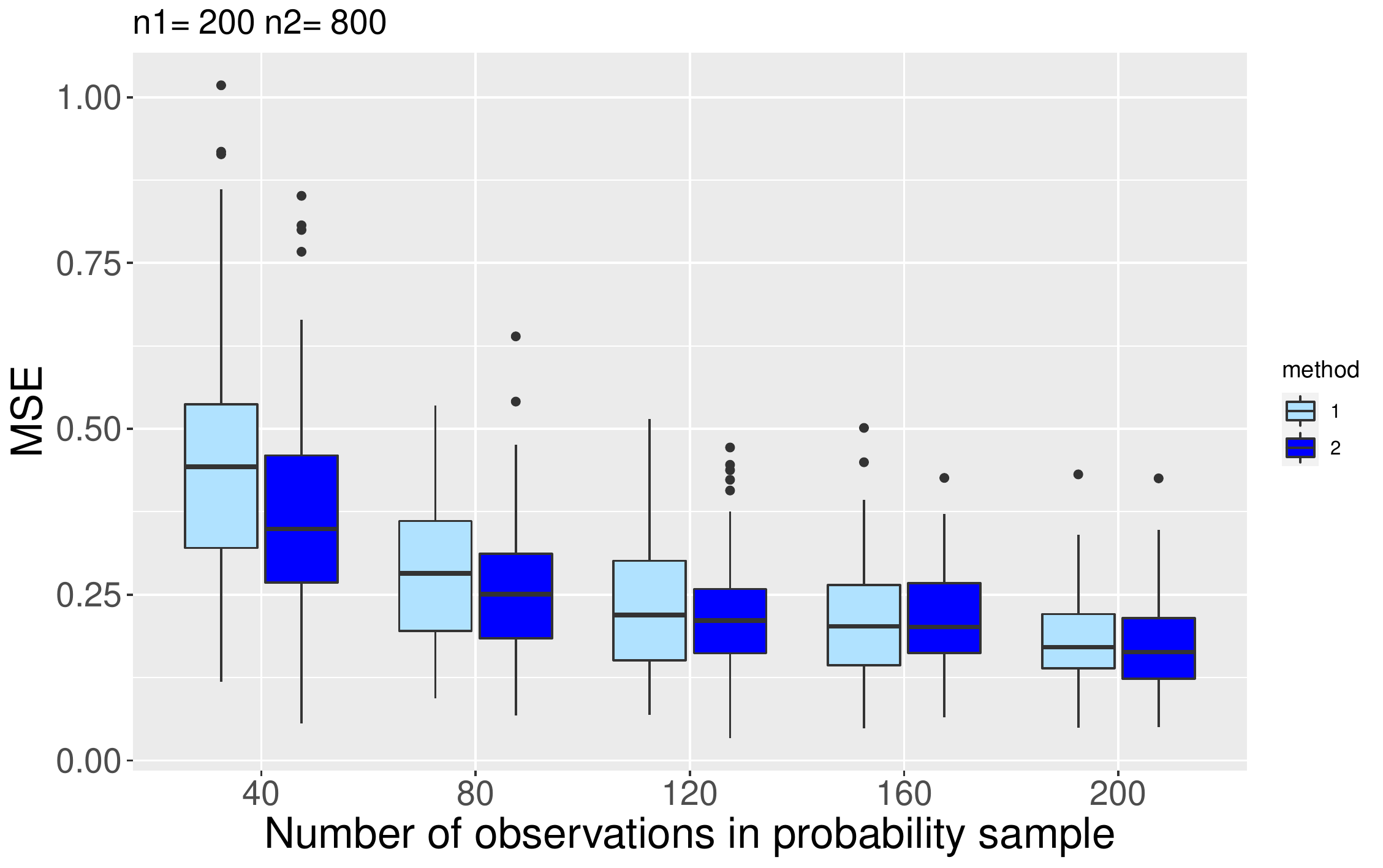}
  \includegraphics[width=0.45\textwidth]{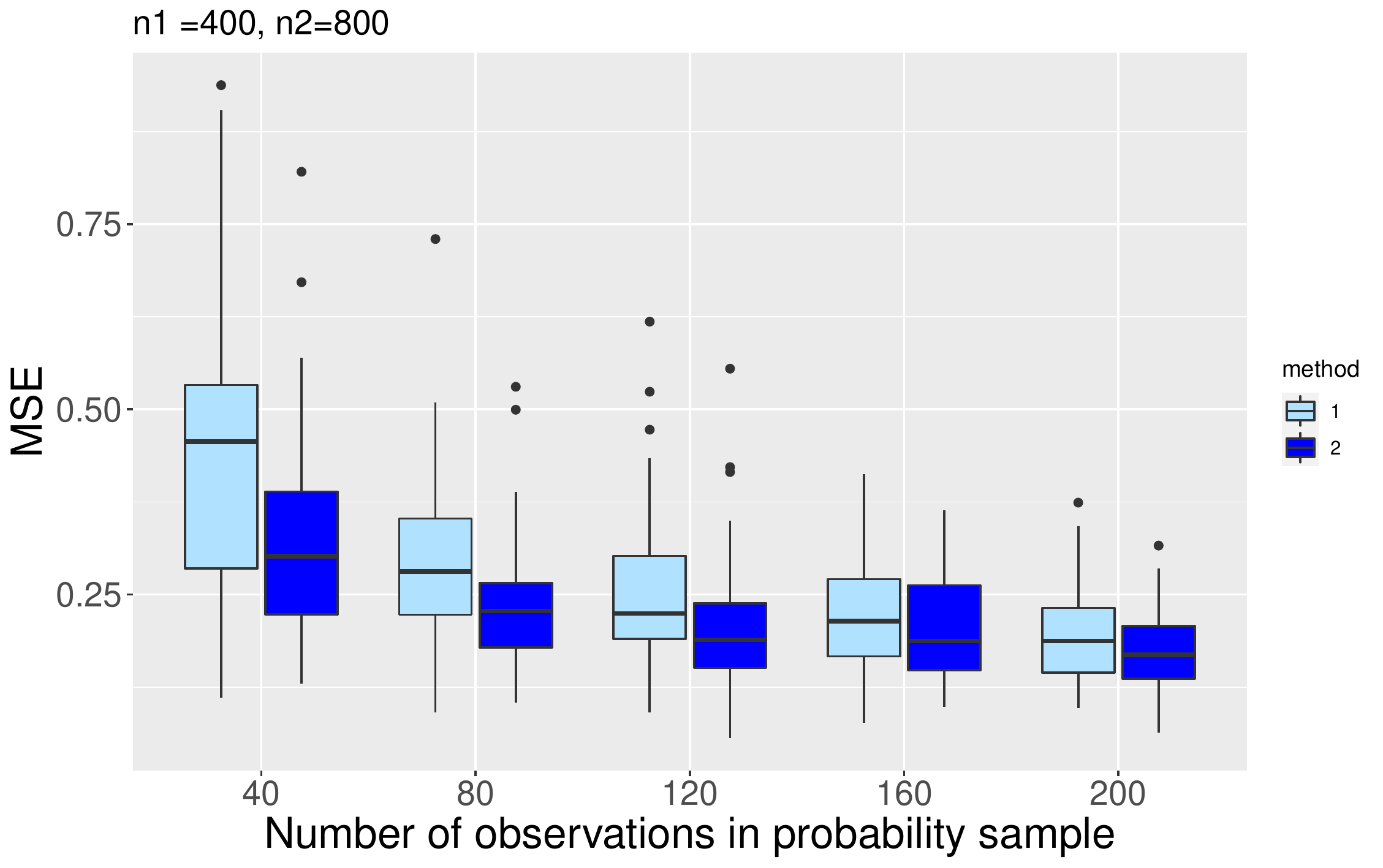}  
	 \caption{MSEs for varying numbers of observations in probability sample, setting 1, method 1 (light) is estimate in probability sample,  method 2 (dark) in extended sample}
  \label{fig:sim11}
\end{figure}

Figure \ref{fig:sim11} shows the mean squared errors of setting 1 with four explanatory variables for varying numbers of observations $n_1,n_2.$ when the number of observations in the probability sample increases from 40 to 200.  For simplicity we used $\alpha_{st}=\alpha_{ch}=0.05$, choice by cross-validation yielded very similar $\alpha$ values and performance typically was comparable. It is seen that the MSEs for both the probability sample estimator (PSE) and the estimator based on the extended sample (ExtE) decrease with increasing sample sizes. The ExtE performs  consistently better than the PSE. Only in the case where there is not much to gain from using the observations in the non-probability sample ($n_1=200$) and there are many polluted observations ($n_2=800$) the dominance becomes weaker, in particular for large probability sample sizes.

\begin{figure}[H]
	\centering
		\includegraphics[width=0.45\textwidth]{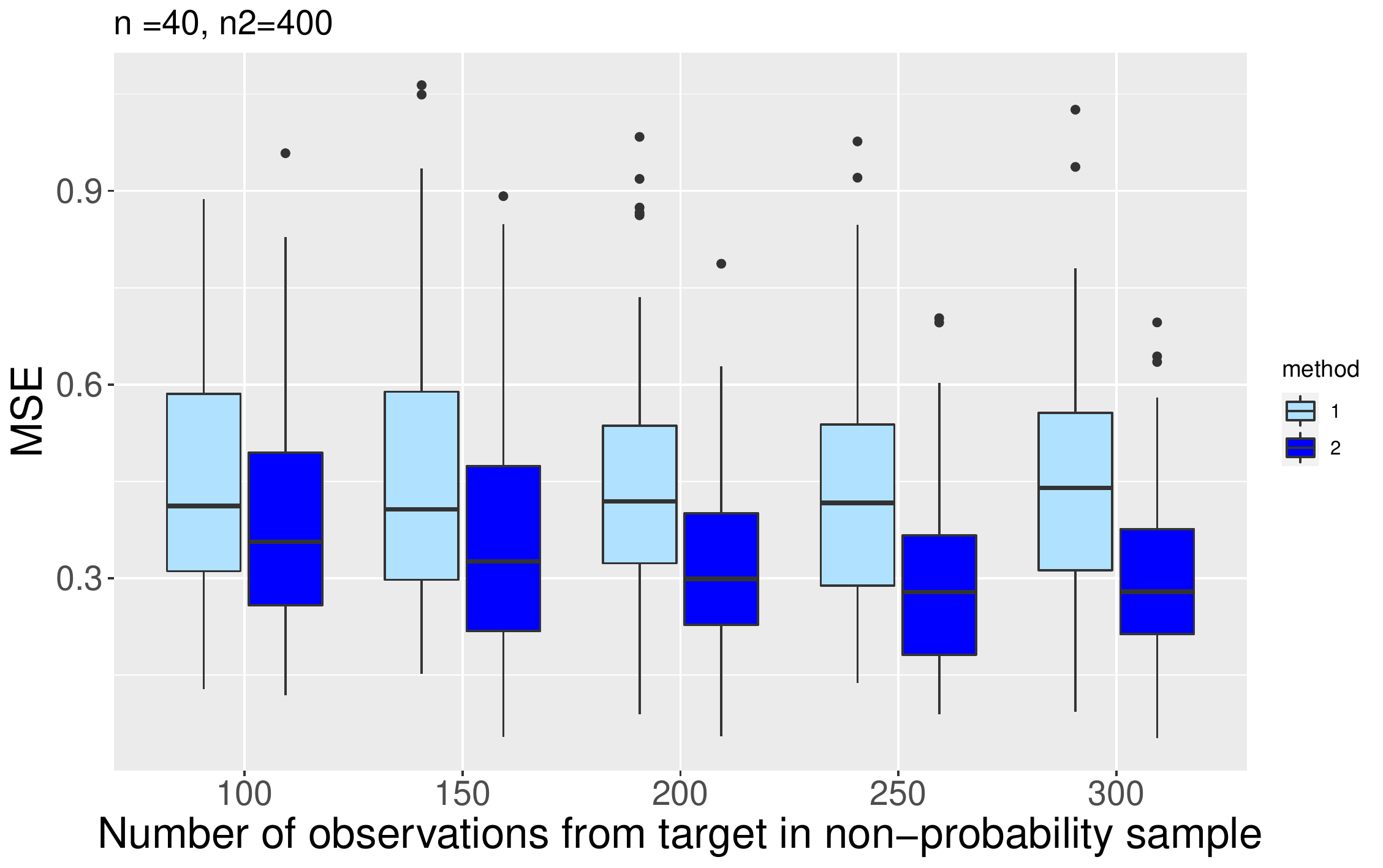}
  \includegraphics[width=0.45\textwidth]{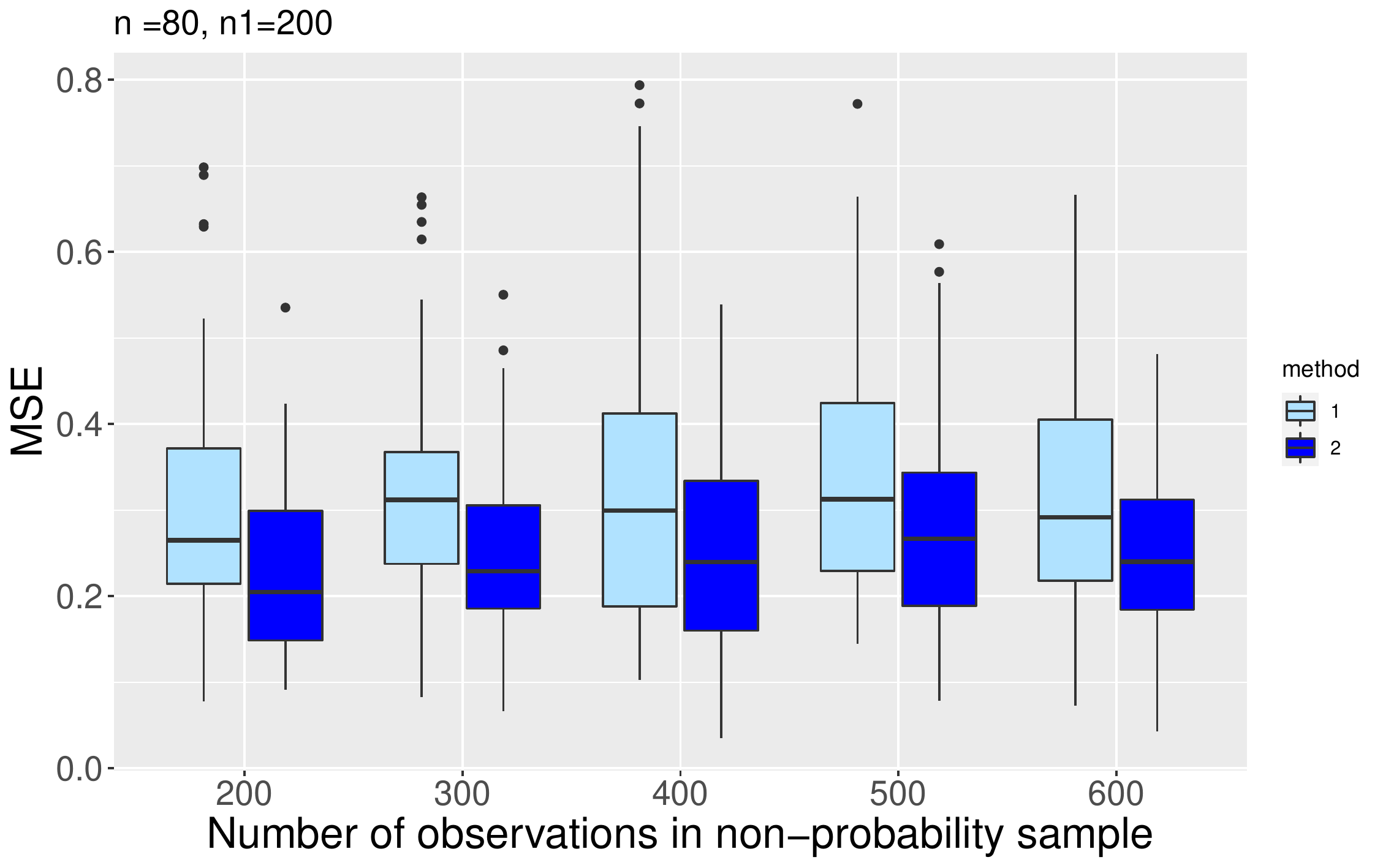}
	 \caption{MSEs setting 1; left: varying numbers of observations from target population in the non-probability sample, right: varying numbers of observations in the polluted data sets,  method 1 is estimate in probability sample,  method 2 in extended sample}
  \label{fig:sim12}
\end{figure}

The left picture in Figure \ref{fig:sim12} shows the comparison of estimates for varying numbers of observations from the target population in the non-probability sample. It is seen that the increased availability of observations from the target population decreases the error in the extended sample. The right picture shows  the effects of varying numbers of observations in the polluted data sets. It is seen that increasing the number of polluted observations hardly affects the performance of the estimates in the extended sample.

\begin{figure}[H]
	\centering
		\includegraphics[width=0.45\textwidth]{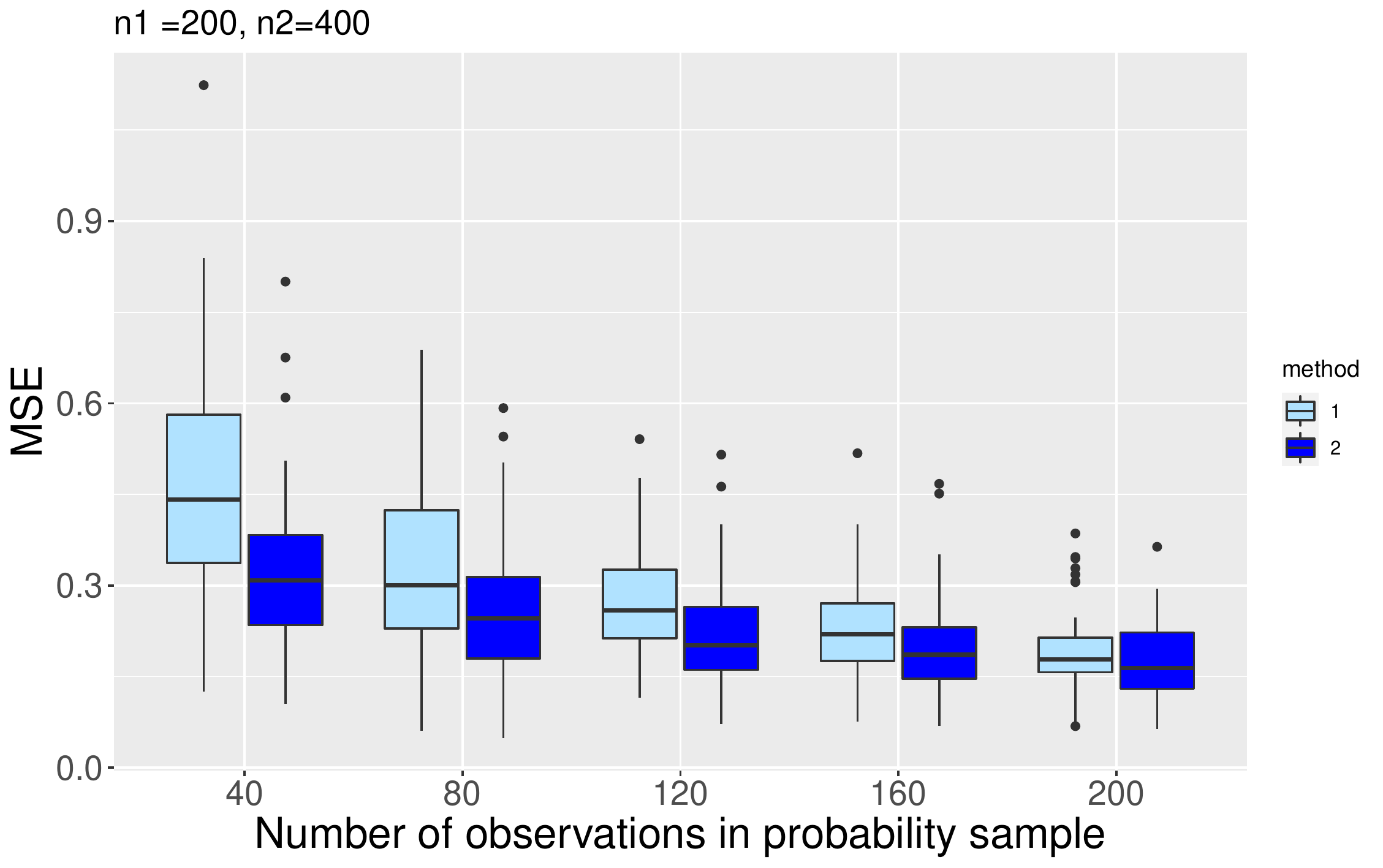}
  \includegraphics[width=0.45\textwidth]{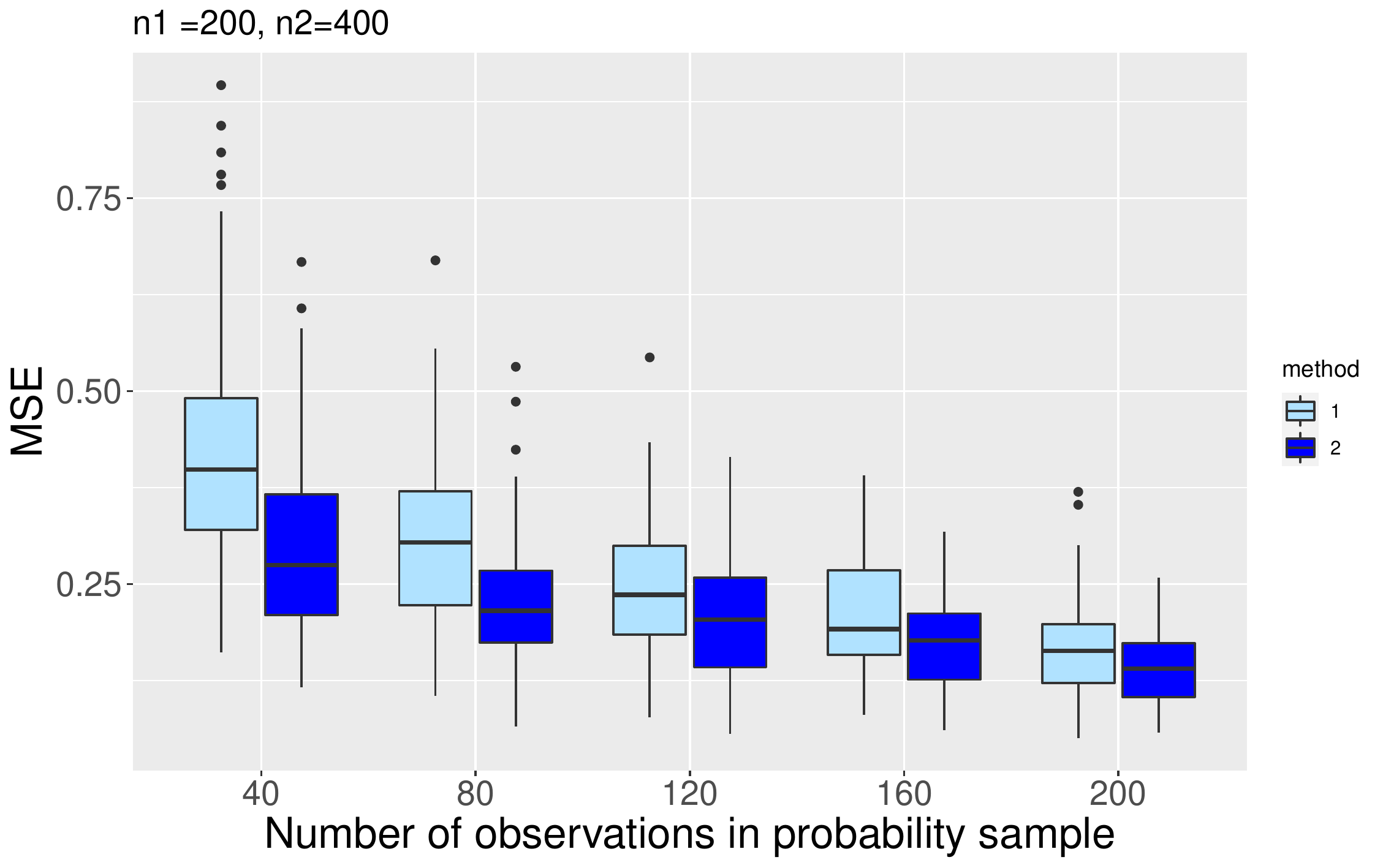}
	 \caption{MSEs for varying numbers of observations in the probability sample, setting 2, left: $\sigma_{par}=2$, right: $\sigma_{par}=4$, method 1 is estimate in probability sample,  method 2 in extended sample}
  \label{sim:2}
\end{figure}

Figure \ref{sim:2} shows the comparisons for varying numbers of observations in probability sample for setting 2 with stronger variation in the polluting sample. The improvement of estimates is similar to that seen in setting 1.

\begin{figure}[H]
	\centering
		\includegraphics[width=0.45\textwidth]{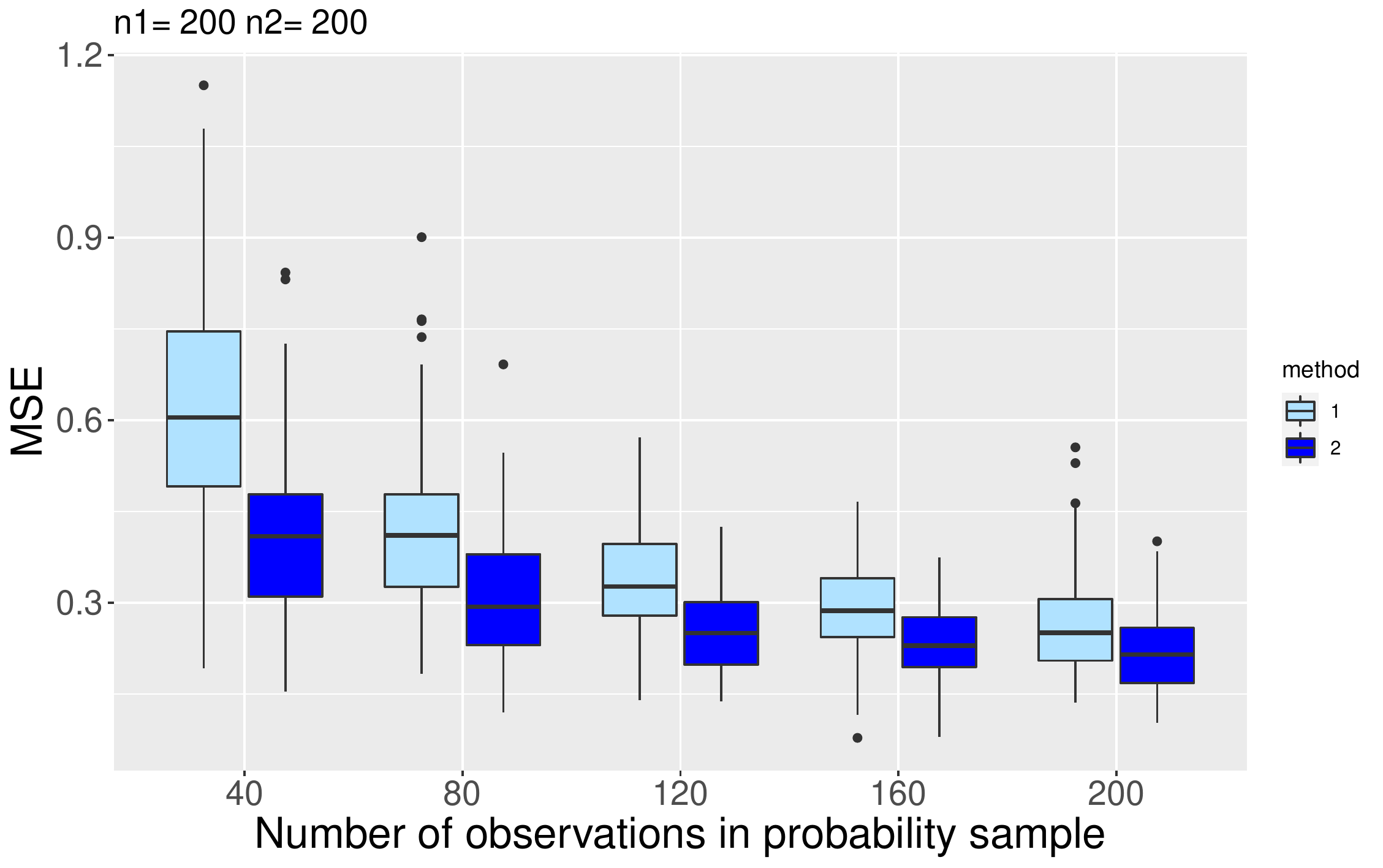}
  \includegraphics[width=0.45\textwidth]{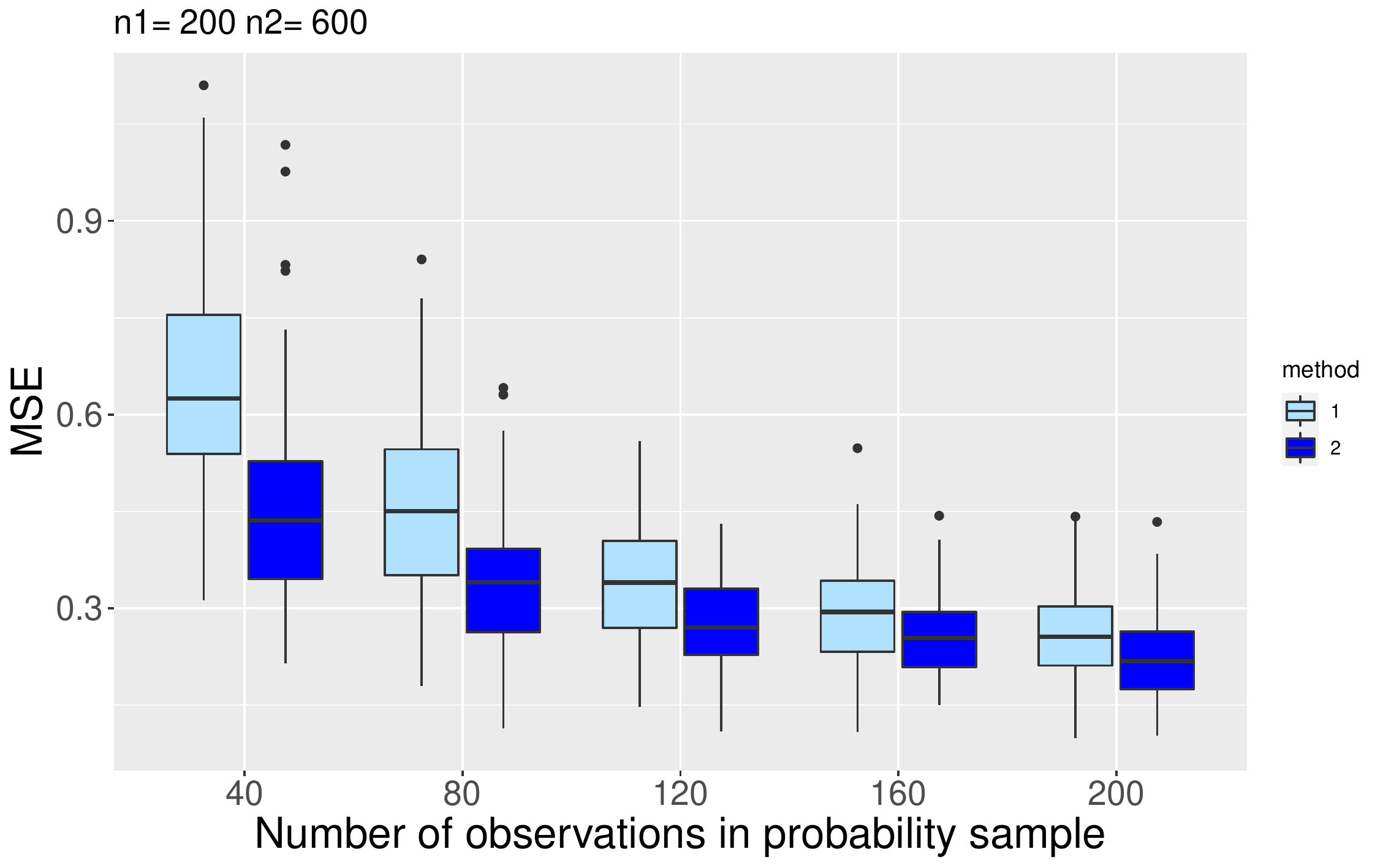}
	 \caption{MSEs for varying numbers of observations in probability sample, setting 3, left: $\sigma_{par}=2$, right: $\sigma_{par}=4$, method 1 is estimate in probability sample,  method 2 in extended sample}
  \label{fig:sim3}
\end{figure}

Figure \ref{fig:sim3} shows the comparisons for varying numbers of observations in probability sample for setting 3 with eight predictors and two settings of observation numbers. The results are very similar to the results seen for four predictors. In particular for small sample sizes of the probability sample the gain in accuracy when using the extended sample is very pronounced.

\section{Standard Errors}\label{sec:stan}

Standard errors for parameter estimates of linear models are easily obtained by using $\cov(\hat \betab)= \sigma^2(\Xb^T\Xb)^{-1}$ and replacing $
\sigma^2$ by the estimator $\hat\sigma^2=\sum_{i}(y_i -\xb^T_i \hat{\betab})^2/(n-(p+1))$. When using the extended sample it is tempting to use the same estimator with the extended design matrix. However, the estimator is severely biased since it ignores the process that augments the data, it simply assumes that the extended data set is a probability sample. The larger number of observations creates the illusion that standard errors are much smaller than they actually are.

A way to obtain reliable estimates is bootstrapping \citep{efron1994introduction, DavHin:97}. For a given data set consisting of the probability sample and the non-probability sample, the whole estimation procedure is carried out repeatedly for data that are obtained by drawing observations from the data set with replacement. The variation of the estimates is used to compute realistic standard errors of parameters.  

An approximation to the true standard errors is obtained by repeating the estimation procedure for $n_{\text{rep}}=100$ drawings for fixed parameters. To obtain bootstrap standard errors, we used $n_{\text{boost}}=100$ bootstrap repetitions. 
We show the results for the following selected simulation settings.

\begin{itemize} 
\item[]Setting B1: Four predictors
 
Probability sample as in setting 1, in particular, $\mu_0^T=\E(\xb)=(1,1,1,1)$, $\betab^T==(1,1,2,3,4)$,

Polluting sample as in setting 1 but $\mu_0^T=\E(\xb)=(2,2,2,2)$, $\betab^T=(1,1,2,3,4)+(1,-1,1,-1)$ are fixed. 


\item[]Setting B2: Same as setting 1 but now stronger variation in probability sample polluting sample  $\var(\varepsilon)=4$ and non-probability sample $\var(\varepsilon)=9$.

\item[]Setting B3: Eight predictors

Probability sample as in setting 3 with $\mu_0^T=(1,\dots,1)$,
regression parameters: $\betab^T=(\beta_{0},\dots,\beta_{8})=(1.0, 0.5, 1.0, 1.5, 2.0, 2.5, 3.0, 3.5, 4.0)$,  $\var(\varepsilon)=1$

Polluting sample: normally distributed,  $\E(\xb)=(3,\dots,3)$,  
regression parameters: $\betab=(1.0, 0.5, 1.0, 1.5, 2.0, 2.5, 3.0, 3.5, 4.0)+(0,1,-1,1,-1,1,-1,1,-1)$,  $\var(\varepsilon)=4$

\end{itemize}

Table \ref{boost1} shows the resulting estimates. \textit{Prob sample} denotes the standard error in the probability sample, \textit{Non-prob naive} are the standard errors obtained as estimates of standard errors when fitting in the extended sample but inoring that the sample has been extended. They are only shown to demonstrate that these estimates should not be used. \textit{Non-prob actual} denotes the approximated actual standard errors and \textit{Non-prob boost} the bootstrap estimates.  It is seen that the nominal standard errors \textit{Non-prob naive}  strongly underestimate the actual standard error.
The boosting errors are comparable to the actual errors but tend to be slightly larger, which makes them somewhat conservative when used in testing. Standard errors in the extended sample are smaller than the standard errors in the probability sample, which typically is also seen in the bootstrap estimates although they are slightly larger than the actual errors.
However, though bootstrap standard errors are not much smaller than standard errors in the probability sample, estimates in the extended sample are more precise as has been demonstrated in the simulations in Section \ref{sec:sim}.

\begin{table}[H]
\caption{Standard errors for selected settings}\label{boost1}
\centering
\begin{tabularsmall}{llllllllllrrrrrrrr}
  \toprule
  & &$\beta_0$ &$\beta_1$ &$\beta_2$ &$\beta_3$ &$\beta_4$ \\
  \midrule
Setting B1  &Prob sample & 0.252 & 0.133 & 0.133 & 0.162 & 0.163  \\
  &Non-prob naive & 0.090  &0.053  &0.052  &0.053  &0.052 \\
  &Non-prob actual & 0.191 &0.102 & 0.111 &0.124 & 0.115  \\
  &Non-prob boost & 0.224  &0.130 &0.143  &0.133  &0.134  \\
\midrule
Setting B2  &Prob sample &0.552 & 0.301 & 0.275 &0.336  &0.289\\
&Non-prob naive & 0.175  &0.100  &0.098  &0.099  &0.099 \\
  &Non-prob actual &  0.368 & 0.199 & 0.202 & 0.213 & 0.192   \\
  &Non-prob boost & 0.436 & 0.267 & 0.275 &0.284 &0.258  \\
  \midrule
Setting B3  &Prob sample & 0.269   & 0.164   & 0.152  & 0.158  &0.166\\
  &Non-prob naive & 0.085 & 0.056  &0.055  &0.055 &0.056\\
  &Non-prob actual & 0.205  & 0.107  &0.106 & 0.109  & 0.115  \\
  &Non-prob boost & 0.257  &0.154  &0.147 &0.152 &0.155  \\

   \bottomrule
\end{tabularsmall}
\end{table}

\section{Illustrative Empirical Studies}\label{sec:emp}

\subsection{Munich rent data}
For illustration we use the Munich rent index data. The variables are 
rent (rent in Euros, metric), 
floor (floor space, metric),
rooms (number of rooms, metric),
district (residential area, nominal). For an extensive description see \citet{fahrmeir2013regression}.
The rent is the dependent variable and floor space and number of rooms are explanatory variables.
The district serves as indicator for the samples. We consider a fixed district, for example ``Inner City'', as the probability sample and observation from other districts as samples from a different source. Part of the sample from other districts could be considered as observations from the probability sample.

Table \ref{tab:rent2} (upper part) shows the estimates for two districts, ``Inner City'' and ``Schwabing'' and the estimates obtained if other districts are used to enlarge the data set. For example, when considering ``Inner City'' as the probability sample one has   47 observations. If one considers ``Bogenhausen'' as the non-probability sample the number of available observations   is 206 (47 from ``Inner City'' and 159 from ``Bogenhausen''). The extended sample size (used observations) when including the observations that  can be considered as coming from the target population  is 198. If the district ``Schwabing'' 
is considered the non-probability sample, the extended sample comprises  203 observations. It is seen that  the standard errors (bootstrap for the enlarged sample) decrease distinctly by enlarging the data sets. 

In the second part of Figure \ref{tab:rent2} the same combinations of probability and non-probability samples are considered but now we do not use the full probability samples to estimate parameters. Since these are real data sets it is unknown which estimates are better, the ones for the probability sample or the ones based on the extended samples. Estimates in the probability sample could suffer from outliers. Therefore, for a fixed probability sample we consider which observations would be included if the same sample is used as non-probability sample. Then we use only the observations that are considered as coming from the probability sample to obtain a reduced probability sample. For the district ``Inner City'' from the original 47 observations only 43 were included in the reduced data set. It is seen that this can distinctly change the estimates. For example, the estimate of floor, which is an important value since it means how much one has to pay per square meter if it is accounted for the number of rooms, is 11.21 in the original sample but 10.46 in the reduced sample, which makes a difference for people renting flats. 

The estimate in the reduced sample can be seen as a robustified estimator since outliers are excluded.   As estimator of the standard error in this case the bootstrap estimator is given since the naive estimator is certainly biased.  Table \ref{tab:rentcross}   
shows how the samples are reduced if outliers are omitted. 'Kept' refers to the number of observations that are kept in the sample and 'deleted' to the number of observations that are left out in the robustified estimator.

\begin{table}[h!]
 \caption{Munich rent data for various districts and enlarged data sets by including neighborhood districts (standard errors obtained by bootstrap).} \label{tab:rent2}
\centering
\begin{tabularsmall}{llccrrrrrrrlllllllllllccccccccc}
  \toprule
&\multicolumn{3}{c}{ } &\multicolumn{2}{c}{ Parameters } &\multicolumn{2}{c}{Std err}\\ 
& &available &used & & & & \\
& & observations& observations&area &rooms &std area &std rooms \\ 
  \midrule
&Inner city &47 &47&11.21       &-92.56 &1.866   &58.072\\
&Enlarged by Bogenhausen &206&198 & 11.22       &-64.87 &1.569& 43.937\\
&Enlarged by Schwabing &212 &203 &10.21       &-53.22  &1.785 &49.880\\
\midrule
&Schwabing &165  &165 &11.16       &-42.36       &1.119 &31.581   \\
&Enlarged  by Maxvorstadt &333  &324 &10.40       &-26.64 &1.203 & 25.907\\
&Enlarged by Bogenhausen &324  &314 &11.27       &-46.03 &1.271 & 24.801\\
\midrule

&Inner city reduced &47  &43&10.46       &-97.99  &2.187& 66.978\\
&enlarged by Bogenhausen  &206  &191 & 10.75       &-69.64 &1.413 &42.952\\
&enlarged by Schwabing  &212 &194 &9.87       &-59.18  &1.524 & 43.958\\
\midrule
&Schwabing reduced &165   &151 &10.11       &-46.08       &2.050 & 63.679   \\
&enlarged by Maxvorstadt  &333    &319 &9.62      &-30.84 &1.083 & 22.314\\
&enlarged by Bogenhausen  &324    &290 &10.35       &-44.74 &1.111 & 22.838 \\

\bottomrule
\end{tabularsmall}
\end{table}

\begin{table}[h!]
 \caption{Munich rent prob and non prob same} \label{tab:rentcross}
\centering
\begin{tabularsmall}{llrrrrrrrrlllllllllllccccccccc}
  \toprule
 &&kept &deleted &proportion kept \\ 
  \midrule
&Inner city&43 &4   &0.914\\
&Schwabing &151 &14&0.915  \\
\bottomrule
\end{tabularsmall}
\end{table}

\subsection{Fears Data}

As a second data set we consider  data from the German Longitudinal Election Study (GLES), which is a long-term study of the German electoral process \citep{GLES}.  The data we are using  originate from the pre-election survey for the German federal  election  in  2017 and consist of responses to various items adressing political fears. The participants were asked: ``How afraid are you due to the use of nuclear energy?'' with answers given on Likert scales from 1 (not afraid at all) to 7 (very afraid). The explanatory variables are age in years, gender (0: male, 1: female) and Abitur (admission certificate to attend university, 1: yes, 0: no)

We consider data collected in a fixed country (in Germany called  Land),  for example Bavaria, as the probability sample and observation from other countries as samples from a different source. Table \ref{tab:GLES1} shows some illustrative examples. First we consider  Bavaria as the target population  and a country that is very close to Bavaria, namely  Baden- Wuertemberg (BW) as additional data set. In the next step we consider data from Baden- Wuertemberg and Hessen as additional
data. In a similar way we consider Berlin as the target population, which is enlraged by using data from the neighbourhood countries Brandenburg (Brand) and Saxony. It is seen that including neighbourhood countries reduces the standard errors, however the effect is less distinct if data from a second country are included.   

Figure  \ref{tab:GLES2} shows fits for the fears data when robustified estimates are used. As before they are obtained by using data from a country as probability sample and the same data as non-probability sample to obtain the reduced sample, which includes only observations that are considered as coming from the target population. Now we excluded   the  variable Abitur, which showed no significant effects  (Table \ref{tab:GLES1}).   
It seems that fits for fixed countries tend to suffer from outliers. The estimated parameters if data from a country are used as probability sample differ distinctly from the fits obtained if in  a first step the data set is reduced. But estimates remain rather constant if data from neighbourhood countries are added to the reduced data set. For Bavaria the standard errors decrease when adding observations but do not improve if Berlin is considered the target population.

\newpage
\begin{table}[H]
 \caption{Fears data for various countries and enlarged data sets by including neighborhood countries (standard errors obtained by bootstrap).} \label{tab:GLES1} 
\centering
\begin{tabularsmall}{llrrcrrrrrrrlllllllllllccccccccc}
  \toprule
&\multicolumn{3}{c}{ } &\multicolumn{3}{c}{ Parameters } &\multicolumn{3}{c}{Std err}\\ 
& &available &used & & & & \\
& & obs& obs  &Age       &Gender       &Abitur & Age       & Gender       & Abitur \\ 
  \midrule
&Bavaria &317 &317 &0.010       &0.566      &-0.294    &0.005  &0.193  &0.209 \\
&enlarged by BW  &522 &507 &0.012   &  0.619      &-0.273     &0.004 &0.166 &0.206 \\
&enlarged by BW, Hessen  &837 &795 &0.014       &0.600      &-0.226     &0.004 &0.165 &0.197 \\
\midrule
&Berlin &62 &62 &0.007       &0.816      &0.148    &0.011  &0.452  &0.450 \\
&enlarged by Brand  &208 &189  &0.011       &0.572       &0.567    &0.010 &0.429 &0.413 \\
&enlarged by Brand, Saxony  &380 &341  &0.014       &0.703       &0.246    &0.010 &0.435 &0.404 \\
\bottomrule
\end{tabularsmall}
\end{table}

\begin{table}[H]
 \caption{Fears data for various countries and enlarged data sets by including neighborhood countries, robustified estimates by initially reducing the probability sample. } \label{tab:GLES2} 
\centering
\begin{tabularsmall}{llrrcrrrrrrrlllllllllllccccccccc}
  \toprule
&\multicolumn{3}{c}{ } &\multicolumn{2}{c}{ Parameters } &\multicolumn{2}{c}{Std err}\\ 
& &available &used & & & & \\
& & obs& obs  &age       &gender        & std age       & std gender        \\ 
  \midrule
&Bavaria all obs&317 &317 &0.012       &0.586           &0.005  &0.193   \\
 \midrule
&Bavaria reduced  &317 &290 &0.020       &0.874           &0.004 &0.171   \\
&enlarged by BW  &495 &464 &0.023      &0.850           &0.003 &0.168  \\
&enlarged by BW, Hessen  &632 &583 &0.024       &0.878      &0.003 &0.166       \\
  \midrule
  \midrule
&Berlin all obs&62 &62 &0.006     &0.821           &0.011  &0.448   \\
 \midrule
&Berlin reduced &62 &55 &0.009     &0.730           &0.009  &0.363 \\
&enlarged by Br  &201 &152 &0.009      &0.801           &0.009 &0.389  \\
&enlarged by Brand, Saxony  &373 &265  &0.014       &0.787        &0.009 &0.410 \\
\bottomrule
\end{tabularsmall}
\end{table}

\section{Concluding Remarks}
It has been demonstrated that the inclusion of selected observations from a different sample can strongly improve parameter estimation in regression modelling.
The model that has been considered was the classical linear model with least squares estimates. The basic methodology, including observations based on the investigation of residuals,  can also be used if alternative estimators are used. In regression with many explanatory variables nowadays shrinkage estimators and other regularized estimators \citep{Tibshirani:96,HasTibFri:2009B} are in common use. It is straightforward to use residuals for these estimators to select observations. The approach could also be used in the fitting of extended models as additive models, generalized additive models \citep{HasTib:90,Wood2017} and generalized linear models \citep{McCNel:83,FahTut:2001} by using appropriate residuals.

The method to enlarge the probability sample used here is an in-or-out procedure. In future research one might alternatively compute weights based on residuals and change in parameters, and put weights on observations in the extended sample. Observations in the probability sample might get weights 1 but observations that are included get smaller weights depending on the values of residual measures.

\bibliography{literatur}

\section*{Appendix }
\subsubsection*{Illustration Univariate Predictor }
Settings (a) and (b) were considered in the text, now we consider one additional setting:

\begin{itemize} 
\item[]Setting (c):
Probability sample:normally distributed,  $\mu_0^T=\E(x)=1$,
regression parameters: $\betab^T=(\beta_{0},\beta_{1})=(1,1)$,  $\var(\varepsilon)=1$

Polluting sample: normally distributed, $\E(x)=2$, 
regression parameters: $\betab^T=(3,-2)$,  $\var(\varepsilon)=4$
\end{itemize}

\begin{figure}[H]
	\centering
		\includegraphics[width=0.45\textwidth]{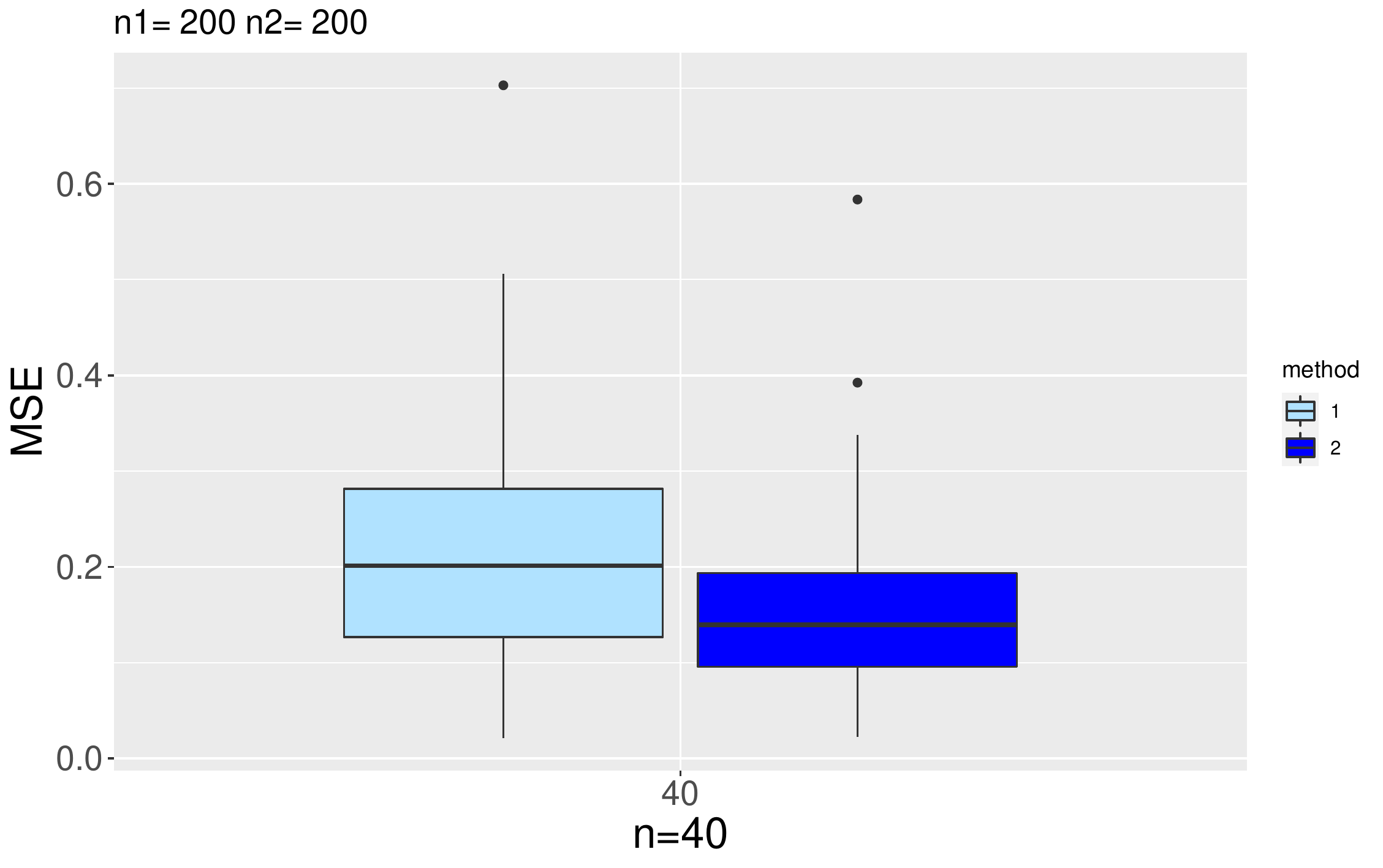}
  \includegraphics[width=0.45\textwidth]{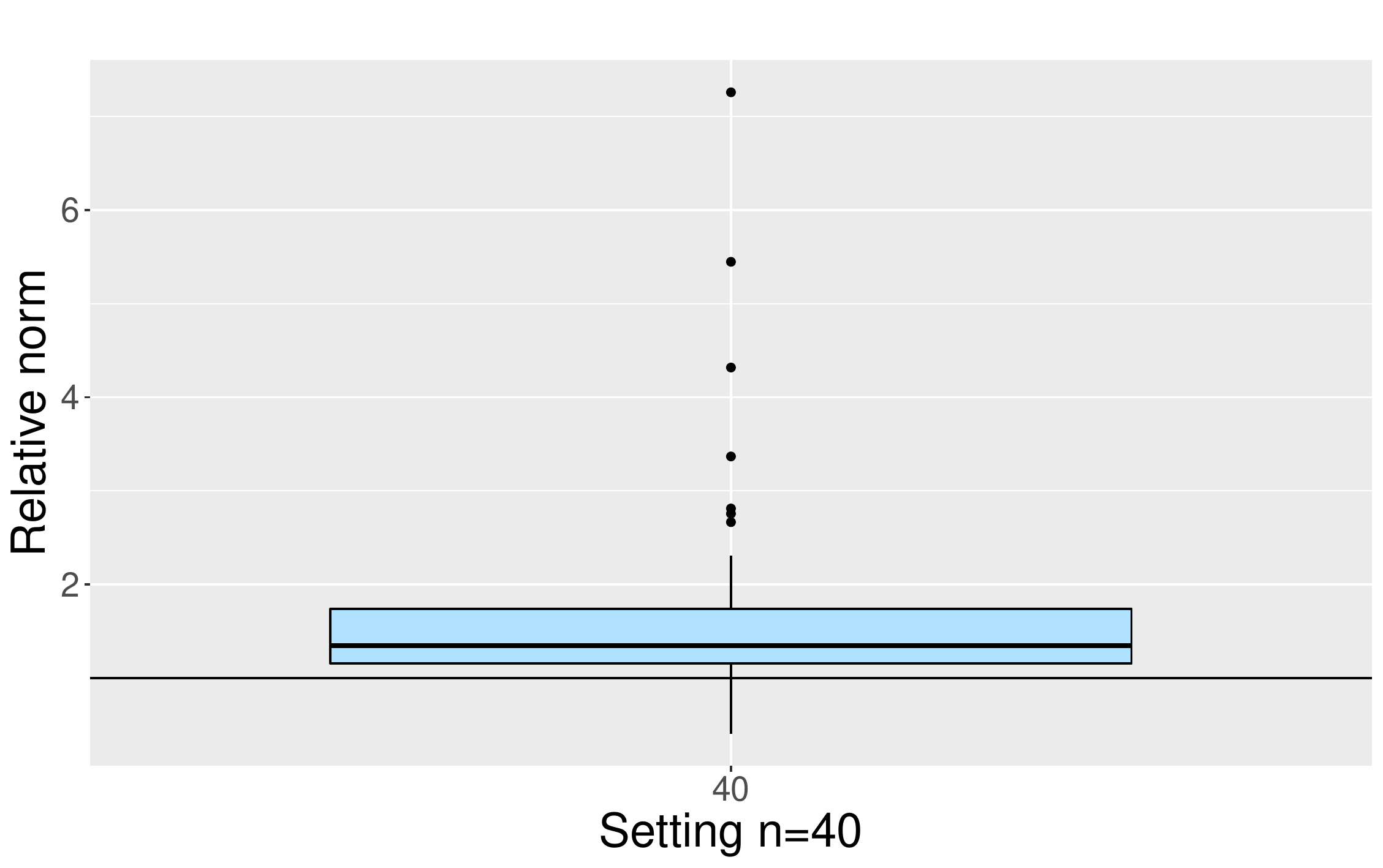}
  \includegraphics[width=0.45\textwidth]{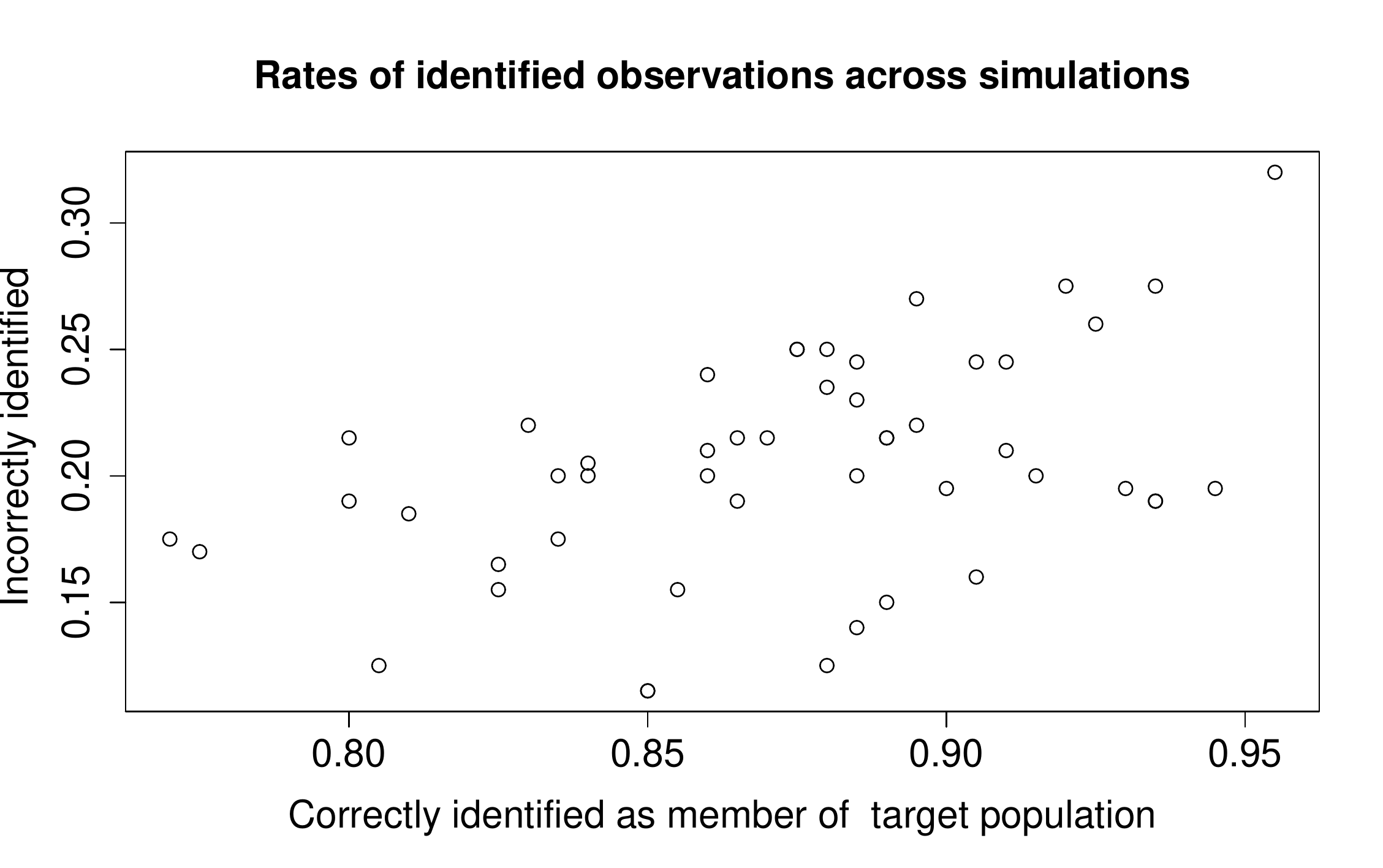}
  \includegraphics[width=0.45\textwidth]{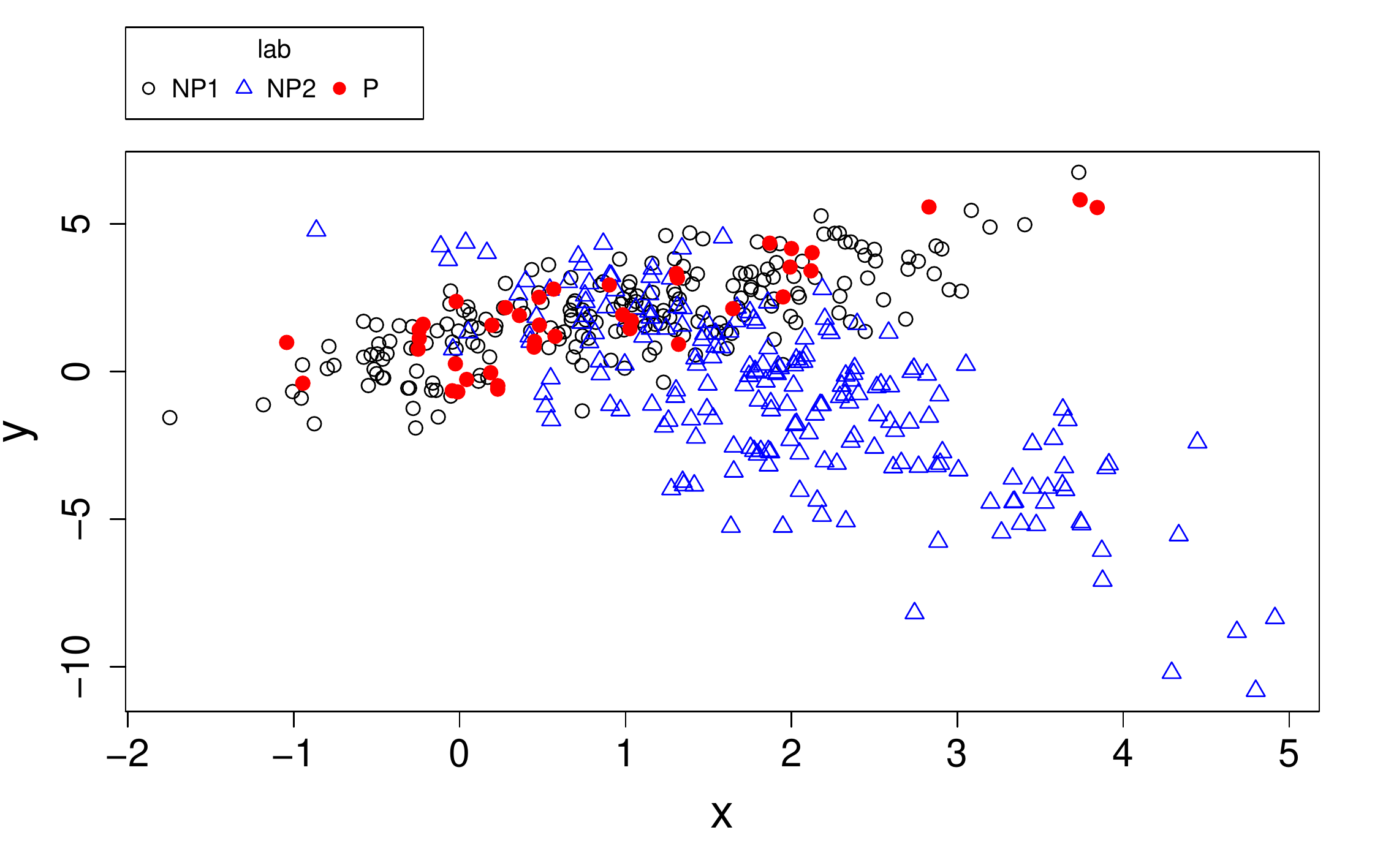}
  \includegraphics[width=0.45\textwidth]{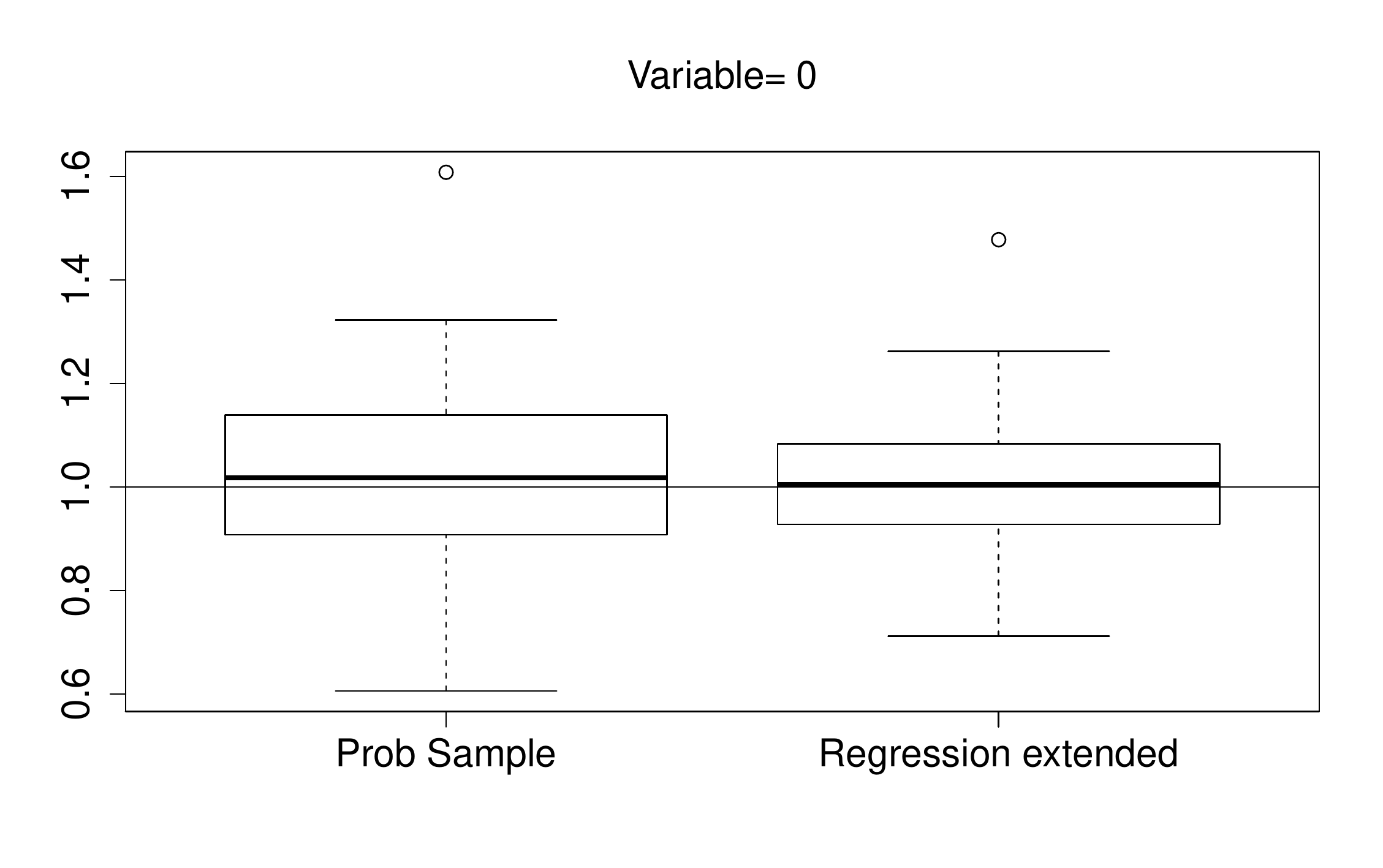}
  \includegraphics[width=0.45\textwidth]{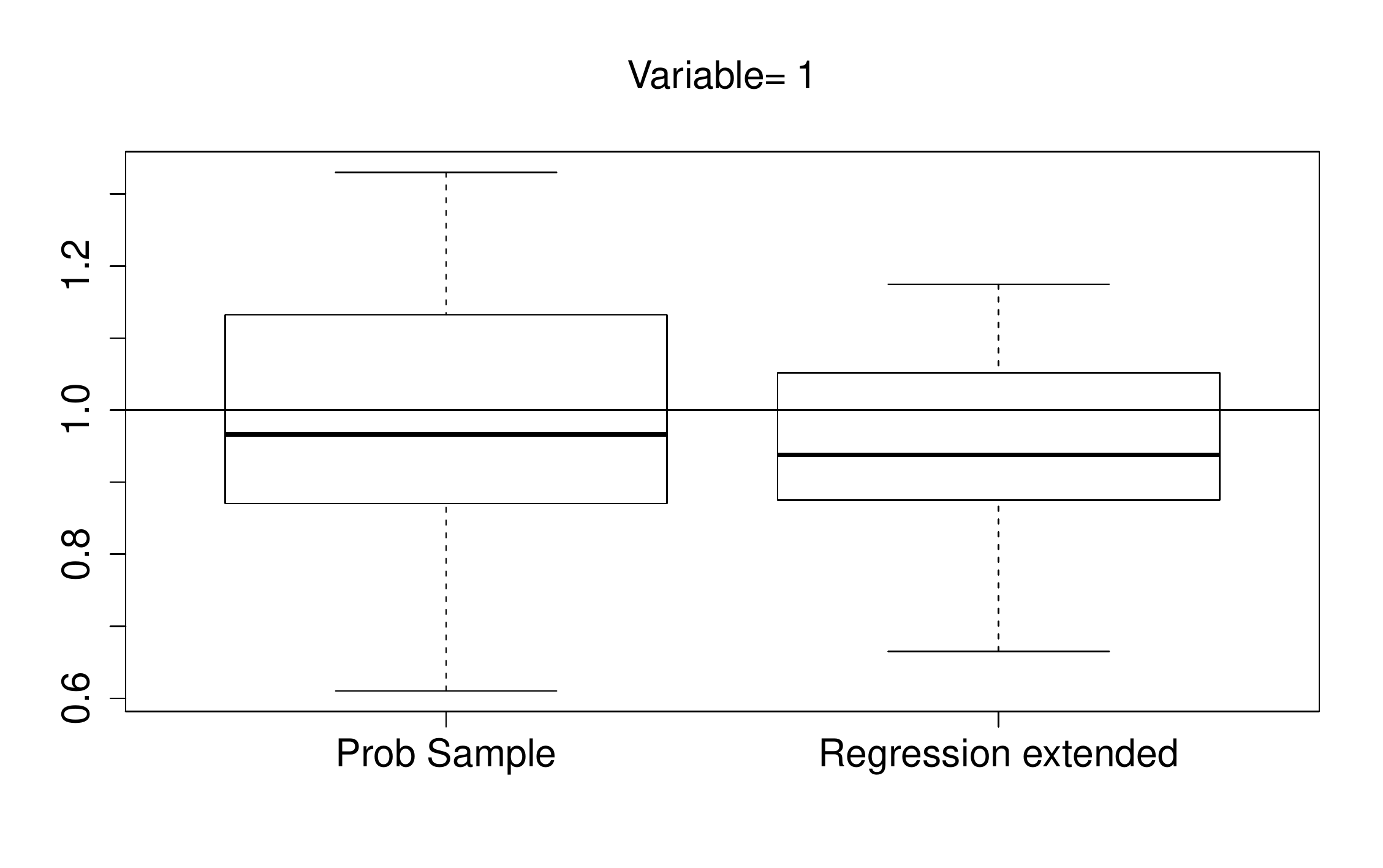}  
	 \caption{Simulation univariate c}
  \label{fig:simc}
\end{figure}

Figure \ref{fig:simc} show the results.

\subsubsection*{Further Simulations Multivariate Predictor }

 We consider again Setting 1 with four predictors but now the noise in the probability sample is larger than in the polluted sample,
 $\var(\varepsilon)= 4$ in probability sample, $\var(\varepsilon)= 1$ in polluted sample. In this case the choice $\alpha_{st}=\alpha_{ch}=0.05$ turns out to perform less convincing. Therefore we used cross-validation to obtain the pictures in \label{fig:simlargercross}. The average values of $\alpha$s ($\alpha_{st}=\alpha_{ch}$) chosen by cross validation for the five settings was $0.155, 0.141, 0.203, 0.190, 0.193$. Choice by cross validation on the full grid of $\alpha$ values from the grid $0.3,0.2,0.1,0.05$  showed no further improvement.

 \begin{figure}[H]
	\centering
		\includegraphics[width=0.45\textwidth]{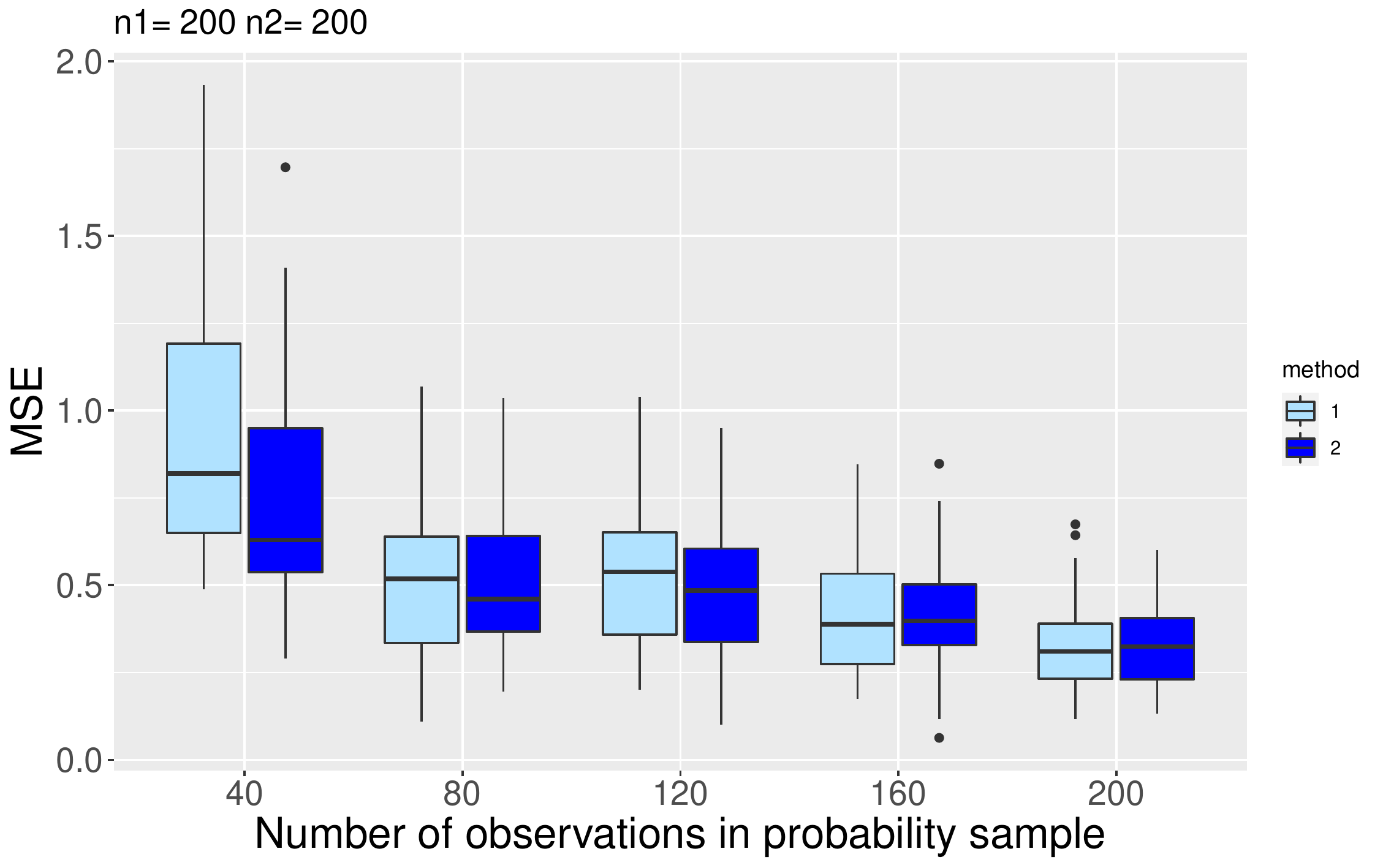}
  \includegraphics[width=0.45\textwidth]{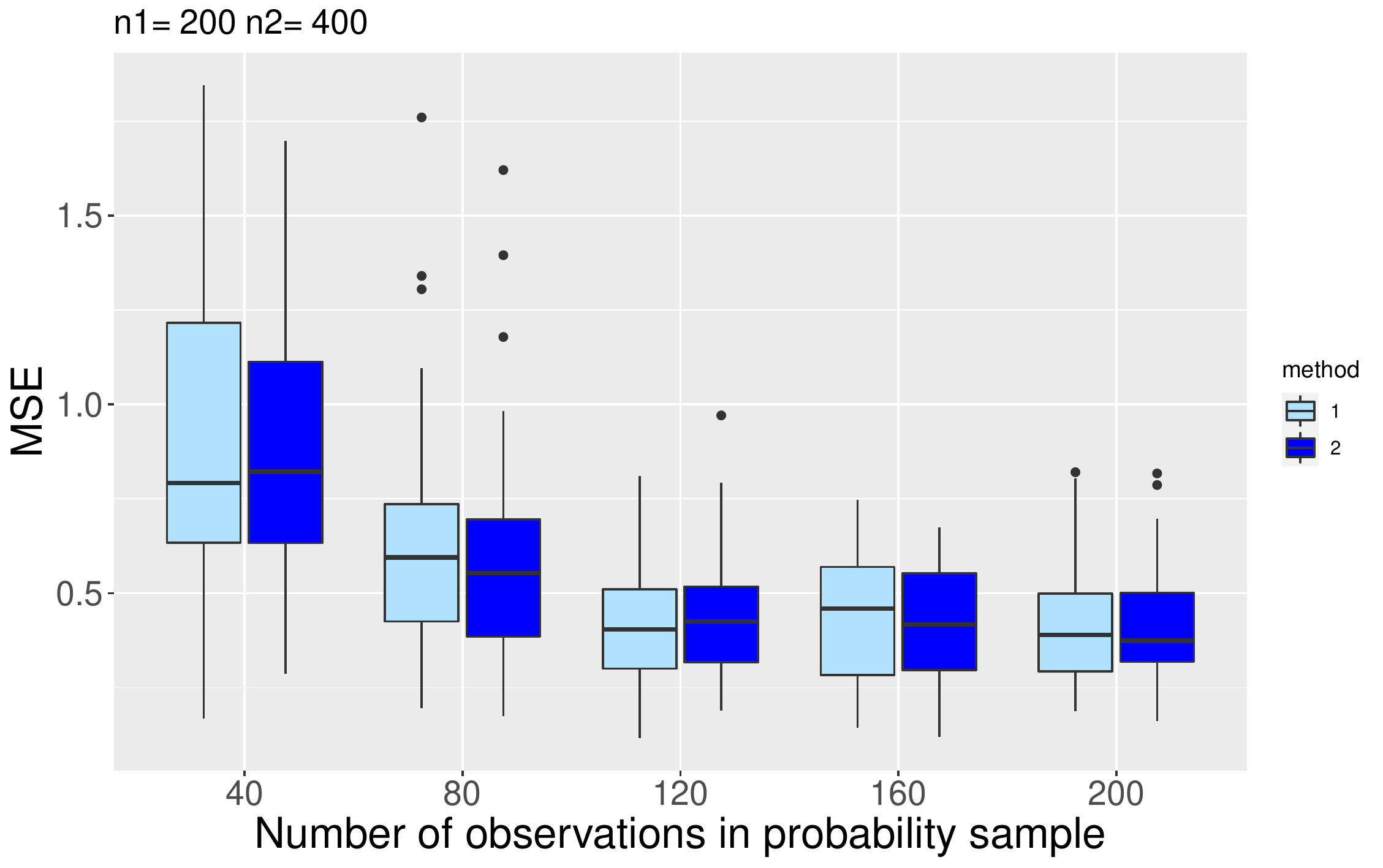}
  \caption{Simulation setting 1 with noise in the probability sample  larger than in the polluted sample, cross-validated.}
  \label{fig:simlargercross}
\end{figure}
\end{document}